\documentclass[12pt]{article}
 \pdfoutput=1

\usepackage{amsmath,amssymb,amsfonts}
\usepackage{hyperref}
\usepackage{graphicx}
\usepackage{color}

\usepackage{mcite}

\newcommand{\fref}[1]{Fig.~\ref{f.#1}}
\newcommand{\eref}[1]{Eq.~(\ref{e.#1})}

\newcommand{\cref}[1]{Chapter~\ref{c.#1}}

\newcommand{\barray}{\begin{eqnarray}}
\newcommand{\earray}{\end{eqnarray}}
\newcommand{\nn}{\nonumber \\}

\newcommand{\beq}{\begin{equation}}
\newcommand{\eeq}{\end{equation}}
\newcommand{\ba}{\begin{array}}
\newcommand{\ea}{\end{array}}
\newcommand{\bea}{\begin{eqnarray}}
\newcommand{\eea}{\end{eqnarray} }
\newcommand{\be}{\begin{eqnarray}}
\newcommand{\ee}{\end{eqnarray} }
\newcommand{\bal}{\begin{align}}
\newcommand{\eal}{\end{align}}
\newcommand{\bi}{\begin{itemize}}
\newcommand{\ei}{\end{itemize}}
\newcommand{\ben}{\begin{enumerate}}
\newcommand{\een}{\end{enumerate}}
\newcommand{\bc}{\begin{center}}
\newcommand{\ec}{\end{center}}
\newcommand{\bt}{\begin{table}}
\newcommand{\et}{\end{table}}
\newcommand{\btb}{\begin{tabular}}
\newcommand{\etb}{\end{tabular}}
\newcommand{\bvec}{\left ( \ba{c}}
\newcommand{\evec}{\ea \right )}

\def\cl{{\mathcal L}}

\def\mev{\, {\rm MeV}}
\def\gev{\, {\rm GeV}}

\def\pa{\partial}

\newcommand{\tr}{\mathrm T \mathrm r}

\newcommand\simlt{\stackrel{<}{{}_\sim}}
\newcommand\simgt{\stackrel{>}{{}_\sim}}

\newcommand{\ti}{\tilde}
\def\hc{{\rm h.c.}}

\newcommand{\eps}{\epsilon}

\numberwithin{equation}{section}

\begin{document}

\begin{titlepage}
\vspace{-1cm}
\begin{flushright}
\small
LPT-ORSAY 12-77
\end{flushright}
\vspace{0.2cm}
\begin{center}
{\Large \bf Higgs After the Discovery: A Status Report}
\vspace*{0.2cm}
\end{center}
\vskip0.2cm

\begin{center}
{\bf  Dean Carmi$^{a}$,  Adam Falkowski$^{b}$, Eric Kuflik$^{a}$,  Tomer Volansky$^{a}$ and Jure Zupan$^{d}$}

\end{center}
\vskip 8pt

\begin{center}
{\it $^{a}$ Raymond and Beverly Sackler School of Physics and Astronomy, Tel-Aviv University, Tel-Aviv 69978, Israel}  \\
{\it $^{b}$ Laboratoire de Physique Th\'eorique d'Orsay, UMR8627--CNRS,\\ Universit\'e Paris--Sud, Orsay, France}\\
{\it $^{d}$ Department of Physics, University of Cincinnati, Cincinnati, Ohio 45221, USA}
\end{center}

\vspace*{0.3cm}

\vglue 0.3truecm

\begin{abstract}
\vskip 3pt \noindent

 Recently, the ATLAS and CMS collaborations have announced the discovery of a 125 GeV particle,
commensurable with the Higgs boson.
We analyze the 2011 and 2012 LHC and Tevatron Higgs data in the context of simplified new physics models, paying close attention to models which can enhance the diphoton rate and allow for a natural weak-scale theory.
Combining the available LHC and Tevatron data in the $h\to ZZ^* \to 4l$, $h\to WW^* \to l\nu l \nu$, $h \to \gamma \gamma$, $h jj \to \gamma \gamma jj$ and $h V \to b \bar b V$ channels, we derive constraints on an effective low-energy theory of the Higgs boson.
We map several simplified scenarios to the effective theory, capturing numerous new physics models such as supersymmetry, composite Higgs, dilaton.
We further study models with extended Higgs sectors which can naturally enhance the diphoton rate.
We find that the current Higgs data are consistent with the Standard Model Higgs boson and, consequently, the parameter space in all models which go beyond the Standard Model is highly constrained.

\end{abstract}

\end{titlepage}

\newpage

\section{Introduction}

ATLAS and CMS \cite{Gianotti:gia12,Incandela:inc12} have just presented an update of the Higgs searches, independently combining  about 5~fb$^{-1}$ of data collected at $\sqrt s = 7$~TeV and more than 5~fb$^{-1}$ at $\sqrt s = 8$~TeV.
Both experiments observe a distinct excess in the diphoton invariant mass spectrum near $125$~GeV with a local  significance of $4.5 \sigma$ and $4.1 \sigma$, respectively.
Moreover, an excess of 4-lepton events with $m_{4l} \simeq 125$ GeV, which can be interpreted as a signal of the $h \to ZZ^* \to 4l$  decay, is observed by both experiments with the significance of $3.4 \sigma$ and $3.2 \sigma$, respectively.
Combining all available channels,  the significance of the signal is around $5.0 \sigma$ for both ATLAS and CMS.
Given these data, the existence of a new resonance near 125 GeV is now established beyond reasonable doubt.
The remaining open question is the precise nature of that resonance. 
{\em Is it a Higgs boson?} {\em Is it the Standard Model Higgs boson?} If not, {\em what sort of new physics is being favored or disfavored by the Higgs data?}

To address these questions, in this paper we combine the available ATLAS \cite{ATLAS_gaga,ATLAS_ZZ4l,ATLAS_WW2l,ATLAS_bb}, CMS \cite{CMS_gaga,CMS_ZZ4l,CMS_WW2l,CMS_bb}, and Tevatron \cite{TEVNPH:2012ab} data in several channels that are currently most sensitive to a 125 GeV Higgs.
We interpret the data as constraints on an effective theory describing general interactions of a light Higgs boson with matter~\cite{Carmi:2012yp,Azatov:2012bz,Giardino:2012ww}.
In this approach, a small number of couplings captures the leading-order Higgs interactions relevant for production and decay processes probed by the LHC and Tevatron.
The Higgs of the Standard Model (SM) corresponds to one point in the parameter space spanned by the effective theory couplings.
Given the event rates measured by experiments and the corresponding errors (assumed to be Gaussian), we can construct the likelihood functions in the parameter space.
This allows us to determine whether the SM Higgs is consistent with the existing  data, and quantify the possible departures from the SM in a general framework with a minimal number of theoretical assumptions.

We also confront the Higgs data with a number of simplified models beyond the SM (for earlier such studies based on 2011 data, see e.g.~\cite{Espinosa:2012qj} and reference therein).
New degrees of freedom coupled to the Higgs and carrying color and/or electric charge may affect the effective couplings of the Higgs to gluons and photons,
while mixing of the Higgs with beyond Standard Model (BSM) scalars may affect the Higgs couplings to the $W$, $Z$ and the SM fermions.
In this context, we discuss which couplings  and mixing angles are allowed by the data.
Unsurprisingly, preferred models feature an enhanced rate in the diphoton channel, as indicated by the data.
We further put a special focus on whether the allowed regions  are consistent with natural theories, where the new degree of freedom cancel the quantum corrections to the Higgs mass induced by the SM top and electroweak bosons.

The paper is organized as follows.  In the next section we define the Higgs low energy effective action and identify the relevant parameters that are being constrained by the present data.   We calculate the contributions to the action from integrating out new physics particles and derive the mapping of the action to the corresponding rates which are measured by the experiments. In Section~\ref{sec:fits}
we discuss the LHC and Tevatron Higgs data  and show the resulting constraints on the parameters of the Higgs effective action.  In Section~\ref{sec:simplified} we then study the constraints on simplified models.
In doing so, we capture several motivated scenarios such as  little Higgs  and supersymmetric models, $W^\prime$,  and dilaton. Section~\ref{sec:extended} is devoted to models with extended Higgs sectors.  We show that simple extensions such as a doublet-singlet or doublet-triplet Higgs sectors, allow for an enhancement in the diphoton rate in agreement with all other constraints.
We conclude in Section~\ref{sec:conclusions}.

\section{Formalism}

We first lay out in some detail our effective theory approach.
We describe interactions of the Higgs boson with matter using an effective Lagrangian where a small number of leading order operators capture the salient features of Higgs phenomenology. Using the effective Lagrangian we derive the relevant production and decay rates as a function of the effective theory couplings. With these relations at hand, one can then construct the coupling-dependent likelihood function for a set of measurements,  allowing for bounds to be placed on these couplings and the best-fit regions to be identified.
We closely follow Ref.~\cite{Carmi:2012yp};  for similar approach  to analyzing the 2011 Higgs data, see \cite{Azatov:2012bz,Giardino:2012ww}.

\subsection{Higgs Effective Action}

We introduce the effective Lagrangian  defined at the scale of $\mu=m_h \simeq 125$~GeV,
\bea
\label{eq:1}
\cl_{eff}  &=  &
c_V  {2 m_W^2  \over v}  h  \,   W_\mu^+ W_\mu^-  +  c_V   {m_Z^2 \over v} h  \,  Z_\mu  Z_\mu  -  c_{b}  {m_b \over v } h \,\bar b b    -  c_{\tau}  {m_\tau \over v } h  \,  \bar \tau \tau  -  c_{c}  {m_c \over v } h \,   \bar c c
    \nonumber  \\&&
 + c_{g}  {\alpha_s \over 12 \pi v} h \, G_{\mu \nu}^a G_{\mu \nu}^a  +  c_{\gamma} { \alpha \over \pi v} h \, A_{\mu \nu} A_{\mu \nu}\,
-  c_{inv}   h \,   \bar \chi \chi  \ .
\eea
This Lagrangian describes the interactions of a light Higgs scalar with matter,  providing a very general and convenient framework for interpreting the current Higgs searches at the LHC and Tevatron.\footnote{A tacit assumption is that we are using the effective Lagrangian to study processes where the Higgs boson is dominantly produced near threshold.
For exclusive processes requiring Higgs produced with a large  boost, $p_{T,h} \gg m_h$, the contribution of higher order operators may be quantitatively important. }
The couplings of the Higgs boson are allowed to take arbitrary values, parametrized by $c_i$.
To be even more general, we also allow for a coupling  to weakly interacting stable particles $\chi$, leading to an invisible Higgs partial width \cite{Giardino:2012ww,Espinosa:2012vu}.
This effective approach harbors very few theoretical assumptions.
One is that of  custodial symmetry, $c_W=c_Z \equiv c_V$  so as to satisfy  the experimental bounds on the $T$-parameter, with $h$ assumed to be a singlet of custodial isospin.
Another theoretical assumption is that the Higgs width is dominated by decays into up to 2 SM particles; more sophisticated BSM scenarios may predict cascade decays into multiple SM particles which would require a separate treatment.
Finally, we assume that the Higgs is a positive-parity scalar; more generally, one could allow for pseudo-scalar interactions.

The top quark has been integrated out in Eq.~\eqref{eq:1} and its effects are included in the effective dimension-5 Higgs couplings parameterized by $c_g$ and $c_\gamma$.
However these 2 couplings can receive additional contributions from integrating out new physics, and therefore are also kept as free parameters.
At the same order one could include the dimension-5 Higgs coupling  to $WW$ and $ZZ$, however their effects can be in most cases neglected in comparison with the contribution proportional to $c_V$.  We therefore  omit them for simplicity. The Lagrangian should be extended by the  dimension-5 coupling to $Z \gamma$, once measurements in this channel become available \cite{Gainer:2011aa}. To describe  the $t \bar t$ associated Higgs production  process, which may be observable in the 14 TeV LHC run, one would not integrate out the top quark.

\subsection{SM and New Physics Contributions}

In the SM, the terms in the first line of Eq.~\eqref{eq:1} arise at tree-level:
\beq
\label{eq:SM1}
c_{V,\rm SM}  =  c_{b,\rm SM} =  c_{\tau,\rm SM} =  c_{c,\rm SM}=1.
\eeq
The values of these couplings may be easily affected by new physics  if the Higgs boson mixes with other scalars, or if the SM fermions and gauge bosons mix with new BSM particles.
As we show below, one may easily construct models where these couplings are either enhanced or suppressed.

We now discuss the possible contributions to the dimension-5 Higgs couplings in the effective Lagrangian \eqref{eq:1}.
Consider a complex scalar  $S$, a Dirac fermion $f$, and a charged vector $\rho_\mu$ (assumed colorless for simplicity) coupled to the Higgs as
\beq\label{eq:partners}
\cl = -  c_s {2 m_s^2 \over v}  h  S^\dagger S  - c_f {m_f \over v} h \bar f f  +  c_\rho {2 m_\rho^2 \over v} h \rho_\mu^\dagger  \rho_\mu.
\eeq
Integrating these particles, the coupling to gluons  and to photons are affected as
\bea
\delta c_g   &=&  {C_2(r_s) \over 2}  c_s A_s(\tau_s) +  2  C_2(r_f) c_f A_f(\tau_f),
\nn
\delta c_\gamma    &=&   { N(r_s) Q_s^2 \over 24}  c_s A_s(\tau_s) + { N(r_f) Q_f^2 \over 6} c_f A_f(\tau_f)  -  {7 Q_\rho^2 \over 8} c_\rho A_v(\tau_\rho),
\eea
where $\delta c_i = c_i - c_{i,\rm SM} $,  $C_2(r)$ is the quadratic Casimir of the color representation, ${\rm Tr} (T^a T^b) = C_2(r) \delta^{ab}$,
and $N(r)$ is the dimension (the number of colors) of the representation $r$.
The  functions $A_i$ describe the 1-loop contributions of scalar, fermion, and vector particles to the triangle decay diagram. They are defined as
\bea
A_s(\tau) &\equiv& \frac{3}{\tau^2} \left [ f(\tau)  - \tau \right ]\,,
\nn
A_f(\tau) &\equiv& \frac{3}{2\tau^2} \left [ (\tau-1)f(\tau)  + \tau \right ]\,,
\nonumber \\
A_v(\tau) &\equiv& \frac{1}{7\tau^2}\left[3(2\tau-1)f(\tau)+3\tau+2\tau^2\right]\,,
\nonumber \\
f(\tau)
&\equiv&  \left\{ \begin{array}{lll}
{\rm arcsin}^2\sqrt{\tau} && \tau \le 1 \\ -\frac{1}{4}\left[\log\frac{1+\sqrt{1-\tau^{-1}}}{1-\sqrt{1-\tau^{-1}}}-i\pi\right]^2 && \tau > 1 \end{array}\right. \,,
\eea
with $\tau_i  = m_h^2/4m_i^2$.   Note that since $f(\tau) \simeq \tau + \tau^2/3$ for $\tau \ll 1$, one finds $A_i(\tau) \simeq 1$ for the contribution of heavy particles ($2 m_i \gg m_h$).

In the SM $c_g$ and $c_\gamma$ arise from integrating out the top quark,  and thus
\beq
\label{eq:SM2}
c_{g,\rm SM} = A_f(\tau_t)  \simeq 1.03\,, \qquad c_{\gamma,\rm SM} =  (2/9)c_{g,\rm SM}  \simeq 0.23\,.
\eeq
The invisible Higgs width in the SM is negligibly small,
\beq
c_{inv, \rm SM} \simeq0\,.
\eeq

\subsection{Partial Decay Widths and Branching Fractions}

With the help of  the effective theory parameters, $c_i$, we can easily write down the partial Higgs decay widths relative to the SM values.
Starting with the decays mediated by the lower-dimensional interactions in the first line of Eq.~\eqref{eq:1}, we have
\beq
\label{eq:h45}
\Gamma_{bb} \simeq |c_b|^2\Gamma_{bb}^{\rm SM} ,
\ \ \Gamma_{\tau \tau}  \simeq  |c_\tau|^2\Gamma_{\tau \tau}^{\rm SM},
  \ \ \Gamma_{WW} \simeq  |c_V|^2\Gamma_{WW}^{\rm SM} ,\ \
  \Gamma_{ZZ} \simeq  |c_V|^2\Gamma_{ZZ}^{\rm SM},
\eeq
where, for $m_h=125$ GeV, the SM widths  are given by
$\Gamma_{bb}^{\rm SM}=2.3 \mev$, $\Gamma_{\tau \tau}^{\rm SM}=0.25\mev$,
$\Gamma_{WW}^{\rm SM}=0.86 \mev$ and $\Gamma_{ZZ}^{\rm SM}=0.1\mev$ \cite{Dittmaier:2012vm}.
Strictly speaking, Eq.~\eqref{eq:h45}  is valid at leading order.
However, higher order diagrams which involve one $c_i$ insertion leave these relations intact.
Thus, Eq.~\eqref{eq:h45}  remains true when higher order QCD corrections are included.
The decays to gluons and photons are slightly more complicated because, apart from the dimension-5 effective coupling proportional to $c_g, c_\gamma$, they receive contribution from the loop of the particles present in Eq.~\eqref{eq:1}.  One has
\bea
\label{eq:ggrate}
\Gamma_{gg}= {|\hat c_g|^2 \over |\hat c_{g,\rm SM}|^2} \Gamma_{g g}^{\rm SM},
\qquad
\Gamma_{\gamma \gamma}= {|\hat c_\gamma|^2 \over   |\hat c_{\gamma,\rm SM}|^2} \Gamma_{\gamma \gamma}^{\rm SM},
\eea
where, keeping the leading 1-loop contribution in each case, one finds,
\bea
\hat{c}_{g} &=&   c_g + c_b A_f(\tau_b) + c_c A_f(\tau_c),
\\
\hat{c}_{\gamma} &=&   c_\gamma  -  {7  c_V  \over 8} A_v(\tau_W) + \frac{c_b}{18} A_f(\tau_b) + \frac{2 c_c }{9}A_f(\tau_c) + \frac{ c_\tau}{6} A_f(\tau_\tau).
\eea
Numerically, for $m_h \simeq 125 \gev$,  $A_v(\tau_W) \simeq  1.19$ and  $A_f(\tau_b) \simeq -0.06 + 0.09 i$,
so that
\beq
\label{eq:chats}
\hat{c}_{g}  \simeq  1.03 c_g - 0.06 c_b, \qquad \qquad \hat{c}_{\gamma}  \simeq  c_\gamma - 1.04 c_V\,.
\eeq
Consequently, $\hat{c}_{g, \rm SM} \approx 0.97$,  $\hat{c}_{g,,\rm SM} \approx -0.81$.
The SM widths for that same mass are $ \Gamma_{gg}^{\rm SM}\simeq 0.34 \mev$ and  $\Gamma_{\gamma \gamma}^{\rm SM}\simeq 0.008 \mev$.

In order to compute the branching fractions in a given channel we need to divide the corresponding partial width by the total width,
\beq
{\rm Br}(h \to i \bar i )  \equiv  {\rm Br}_{ii}= \frac{\Gamma_{ii}}{\Gamma_{tot}}\,.
\eeq
The latter includes the sum of the width in the visible channels and the invisible width which, for $m_h=125$ GeV, is  $\Gamma_{inv}  \simeq 1.2\times 10^3 |c_{inv}|^2  \Gamma_{tot}^{\rm SM}$.
We can write it as
\beq
\label{eq:ctotdef}
\Gamma_{tot}=  |C_{tot}|^2 \Gamma_{tot}^{\rm SM}\,,
\eeq
  where, for $m_h=125$ GeV,  $\Gamma_{tot}^{\rm SM}\simeq 4.0 \mev$, and
\bea
\label{e.ctot}
|C_{tot}|^2 &\simeq& |c_b|^2  {\rm Br}_{bb}^{\rm SM} + |c_V|^2 \left({\rm Br}_{WW}^{\rm SM}+{\rm Br}_{ZZ}^{\rm SM}\right)+ \frac{|\hat c_g|^2}{\left|\hat c_g^{\rm SM}\right|^2} {\rm Br}_{gg}^{\rm SM} +   |c_\tau|^2 {\rm Br}_{\tau\tau}^{\rm SM} +  |c_c|^2 {\rm Br}_{cc}^{\rm SM} +  { \Gamma_{inv} \over   \Gamma_{tot}^{\rm SM} }
\nonumber \\
&\simeq& 0.58 |c_b|^2+0.24 |c_V|^2 + 0.09  \frac{|\hat c_g|^2}{\left|\hat c_g^{\rm SM}\right|^2}  +  0.06  |c_\tau|^2  +  0.03 |c_c|^2 +{ \Gamma_{inv} \over   \Gamma_{tot}^{\rm SM} } .
\nonumber
\eea
Typically, the total width is dominated by the decay to $b$-quarks and $\Gamma_{tot} \sim c_b^2$, however this scaling may not be valid if the Higgs couples more weakly to  bottoms ($c_b \simlt 0.7$), more strongly to  gauge fields ($c_V \simgt 1.4$), or if it has a significant invisible width ($c_{inv} \simgt 0.03$).

\subsection{Production Cross Sections}
\label{s.hpcs}

Much like the decay rates, one can express the relative cross sections for the Higgs production  processes in terms of the parameters $c_i$.
For the LHC and the Tevatron the currently relevant partonic processes are
\bi
\item Gluon fusion (ggF), $g g \to h $+jets,
\item Vector boson fusion (VBF),  $q q \to h qq$+jets,
\item Vector boson associate production (VH), $q \bar q \to h V$+jets
\ei
The relative cross sections in these channels can be approximated at the leading order  by,
\beq
\label{eq:prod}
{\sigma_{ggF} \over \sigma_{ggF}^{\rm SM}} \simeq {|\hat c_g|^2 \over |\hat c_{g,\rm SM}|^2 },
\qquad
{\sigma_{VBF} \over \sigma_{VBF}^{\rm SM}} \simeq |c_V|^2,
\qquad
{\sigma_{VH} \over \sigma_{VH}^{\rm SM}} \simeq |c_V|^2.
\eeq
Using Eq.~\eqref{eq:prod}, we find that the total inclusive $pp \to h$ cross section $\sigma_{tot}$,
\beq
\label{eq:sigtot}
{\sigma_{tot}  \over \sigma_{tot}^{\rm SM} } \simeq {|\hat c_g|^2  \sigma_{ggF}^{\rm SM} /|\hat c_{g,\rm SM}|^2 + |c_V|^2  \sigma_{VBF}^{\rm SM} + |c_V|^2 \sigma_{VH}^{\rm SM} \over
 \sigma_{ggF}^{\rm SM} +  \sigma_{VBF}^{\rm SM} +  \sigma_{VH}^{\rm SM}} \,,
\eeq
is typically dominated by the gluon fusion process, and therefore it scales as $\sigma_{tot} \sim c_g^2$.

\subsection{Event Rates at the LHC and Tevatron}
\label{s.rates}

The event count in a given channel depends on the product of the Higgs branching fractions, the production cross section, and the selection efficiency.
Experiments present the Higgs results as constraints on $R$ (sometimes denoted as $\mu$ or $\hat \mu$) defined as the event rates relative to the rate predicted by the SM.
These rates  can be easily expressed in terms of the parameters of the effective Lagrangian in Eq. \eqref{eq:1}.

We first discuss the {\em inclusive} ATLAS and CMS searches in the $\gamma \gamma$, $Z Z^*$  and $WW^*$ final states.
In those cases, the selection efficiency is similar for each Higgs production channels, to a good approximation.
Thus, these searches constrain the inclusive rates $R^{\rm incl}$ defined as
\begin{eqnarray}
\label{eq:4}
R_{VV^*}^{\rm incl} &\equiv& {\sigma_{tot}  \over \sigma_{tot}^{SM} }{ {\rm Br}(h \to VV^*) \over {\rm Br}_{SM}(h \to VV^*) }
\simeq  \left|\frac{\hat c_g c_V}{\hat c_{g,\rm SM} C_{tot}} \right|^2,
\nn
R_{\gamma \gamma}^{\rm incl} &\equiv& {\sigma_{tot} \over \sigma_{tot}^{SM} }{ {\rm Br}(h \to \gamma \gamma ) \over  {\rm Br}_{SM}(h \to \gamma \gamma) }
\simeq    \left | {\hat c_g  \hat c_{\gamma} \over  \hat c_{g,SM} \hat c_{\gamma,SM} C_{tot}}\right|^2. \label{eq.Rgg}
\eea
The approximations above hold assuming the Higgs production remains dominated by the gluon fusion subprocess.
In our fits we use  more precise expressions including the contribution of all leading production processes listed in Sec.~\ref{s.hpcs}.

Another category  is  {\em associated production} search channels targeting a particular Higgs production mode.
One important example are the  ATLAS and CMS studies in the $\gamma \gamma+2$ jets final state where kinematic cuts on the jets  were employed to enhance the VBF contribution~\cite{ATLAS_gaga,CMS_gaga}.
In the following we refer to these search channels as the {\em dijet $\gamma \gamma$ category}.
Taking into account the selection efficiencies $\eps_i$ for different production channels, the dijet category searches constrain  the  rate $R_{\gamma \gamma}^{\rm dijet}$ defined as
\beq
R_{\gamma \gamma}^{\rm dijet} \simeq {\eps_{ggF} |\hat c_g|^2 \sigma_{ggF}^{\rm SM} / |\hat c_{g,\rm SM}|^2 +  \eps_{VBF}  |c_V|^2  \sigma_{VBF}^{\rm SM} +  \eps_{VH}  |c_V|^2 \sigma_{VH}^{\rm SM}
\over
\eps_{ggF}  \sigma_{ggF}^{\rm SM}  +  \eps_{VBF} \sigma_{VBF}^{\rm SM}  +  \eps_{VH}\sigma_{VH}} { {\rm Br}(h \to \gamma \gamma ) \over  {\rm Br}_{SM}(h \to \gamma \gamma) }\,,
\eeq
where we used the approximated values given in Eq.~\eqref{eq:SM1}.

Finally, the searches in the $b \bar b$  final  state at the Tevatron and the LHC target the Higgs produced in association with a (leptonically decaying) $W$ or $Z$ boson,
therefore they constrain   $R_{b b}^{\rm VH}$ defined as
\beq
\label{eq:last}
R_{b b}^{\rm VH} \equiv {\sigma (p  p \to V h)  \over \sigma_{\rm SM} (p  p \to V h) }{ {\rm Br}(h \to b \bar b) \over {\rm Br}_{SM}(h \to b \bar b) }
\simeq  \left|\frac{c_V c_b}{C_{tot}} \right |^2.
\eeq

In summary, using Eqs.~(\ref{eq:h45})-(\ref{eq:last}) we can express the observable event rates $R_{ii}$ in terms of the parameters of the effective theory defined by the Lagrangian \eqref{eq:1}.
In the following we use the latest experimental determinations of $R_{ii}$ with the corresponding errors (assumed to be Gaussian) to identify the preferred regions of the parameter space.

\section{Combination and Global Fits}
\label{sec:fits}

\begin{figure}
\bc
\includegraphics[width=0.7\textwidth]{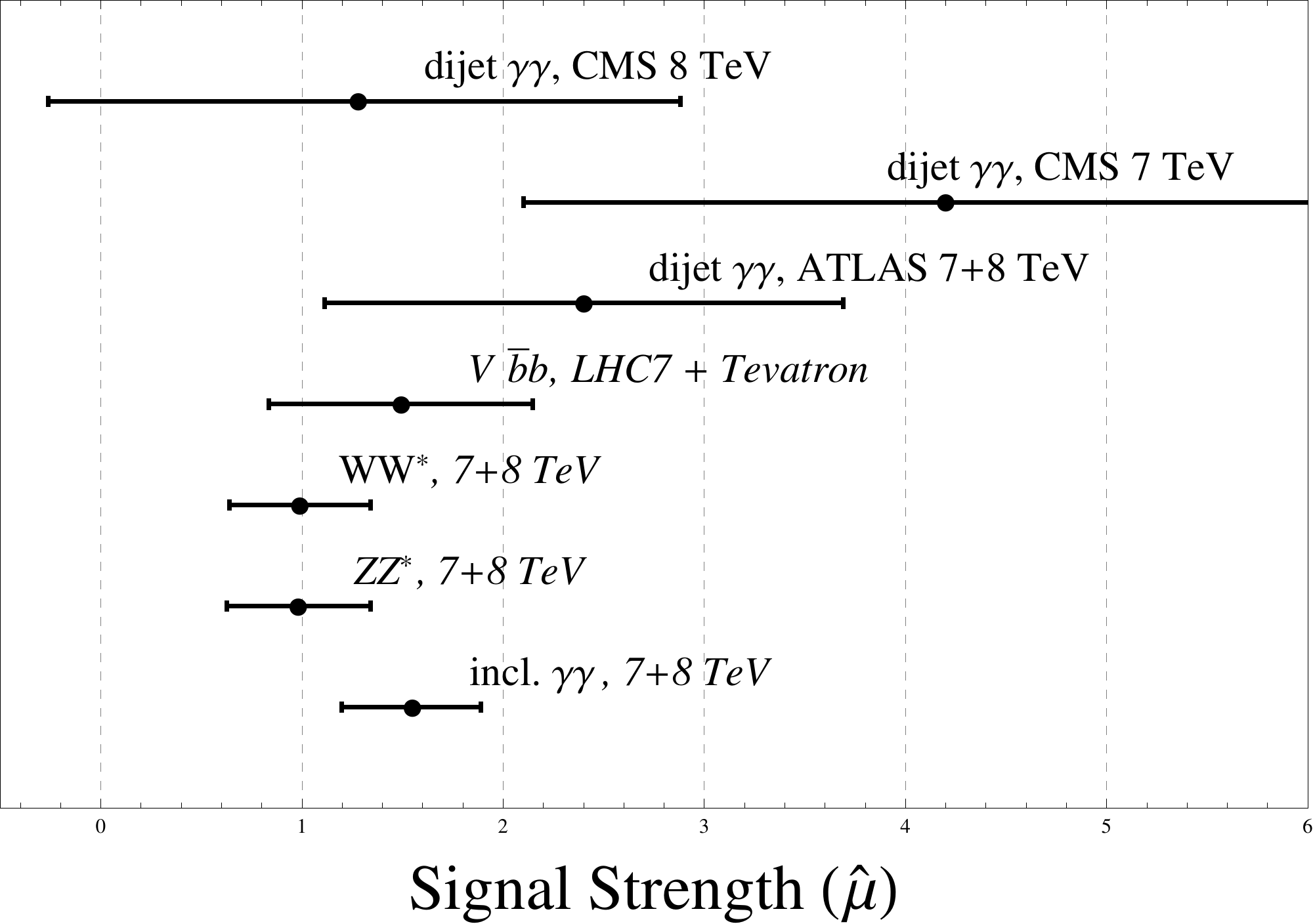}
\ec
\caption{\small The combined signal strength $\hat \mu_{ii} \equiv R_{ii}$ and the corresponding error  for the Higgs search channels used in our analysis.
 \label{f.data}}
\end{figure}

Recently, the LHC  updated the Higgs searches adding almost 6 fb$^{-1}$ per experiment, collected at $\sqrt{s} = 8$~TeV.   Meanwhile, the Tevatron presented a new refined analysis of the full 10~fb$^{-1}$ data set.
Here we focus on the following channels: the inclusive and the dijet tag $h\to \gamma \gamma$ \cite{ATLAS_gaga,CMS_gaga},  $h \to ZZ^* \to 4l$ \cite{ATLAS_ZZ4l,CMS_ZZ4l},  $h \to WW^* \to 2l 2\nu$ \cite{ATLAS_WW2l,CMS_WW2l},  and $Vh \to V b \bar  b$  \cite{ATLAS_bb,CMS_bb,TEVNPH:2012ab} channels.  These are currently the most sensitive search channels for $m_h \simeq 125$~GeV.
The combined central values for the signal strengths $R_{ii}$ in these channels and the corresponding errors  are displayed in \fref{data}.
In the following sections we use these bounds on $R_{ii}$ to constrain  the parameters of the effective theory defined by \eqref{eq:1}. Throughout we assume $m_h= 125$~GeV and Gaussian statistics, and we do not take into account systematic effects which may be significant in the $ZZ^*$ channel and the dijet tag channel of the diphoton analysis.

\begin{figure}[!]
\vspace{0cm}
\bc
\includegraphics[width=0.9\textwidth]{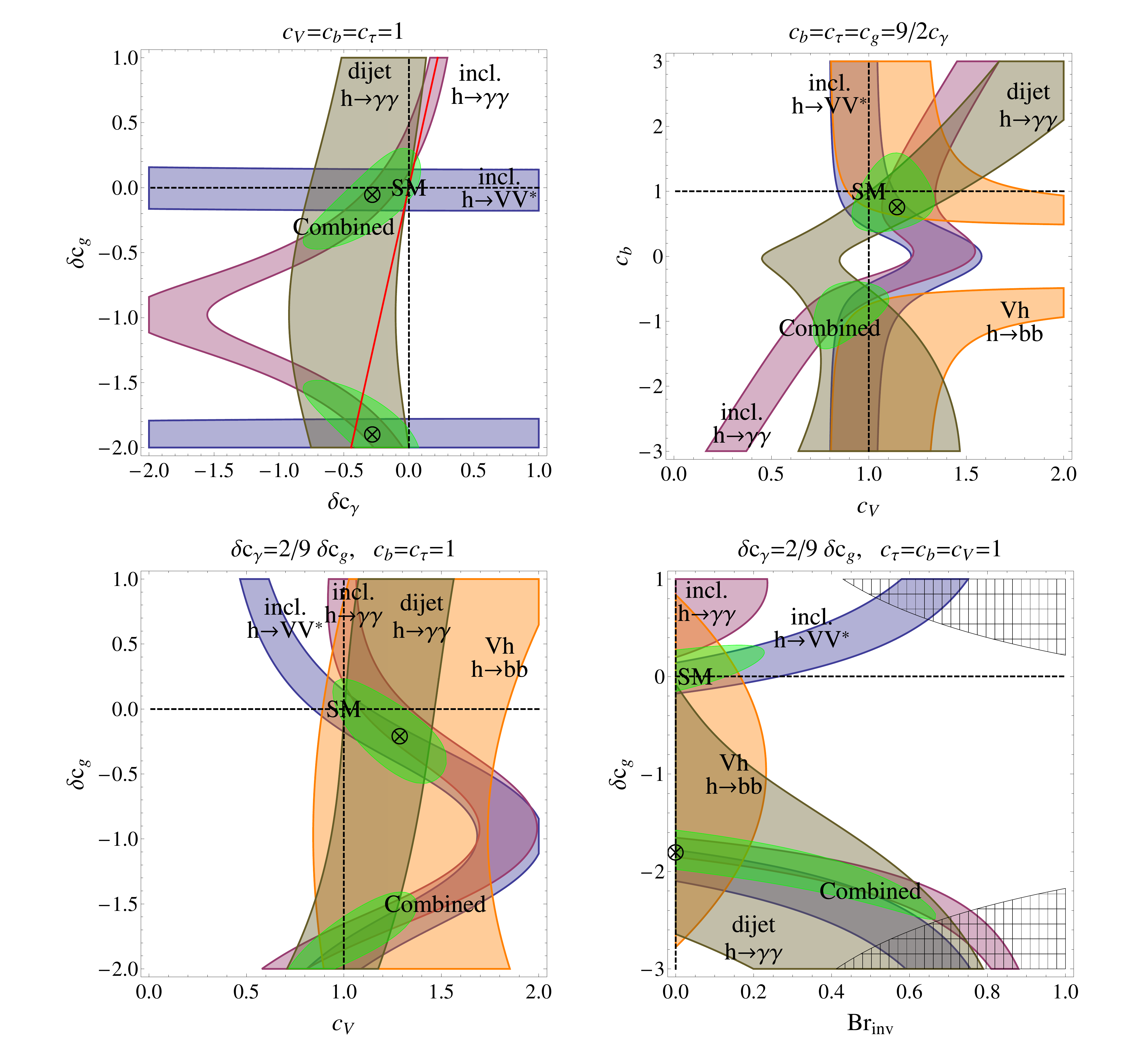}
\ec
\vspace{-1cm}
 \caption{\small
The allowed parameter space of the effective theory given in Eq.~\eqref{eq:1}, derived from the LHC and Tevatron constraints for $m_h = 125$ GeV.
We display the $1\sigma$ allowed regions for the rates in  Eqs.~\eqref{eq:4}-\eqref{eq:last}: $R_{\gamma \gamma}$ (purple), $R_{ZZ}$ (blue), $R_{WW}$ (light grey), $R_{\gamma \gamma jj}$ (beige), and $R_{b \bar b}$ (orange). The ``Combined'' region (green) shows the 95\% CL preferred region arising from all channels. The crossing of the dashed lines is the SM point. The $\bigotimes$ corresponds to the best fit point. The {\bf top-left} plot characterizes models in which loops containing beyond the SM fields contribute to the effective 5-dimensional $h \, G_{\mu \nu}^a G_{\mu \nu}^a$ and $h \, A_{\mu \nu} A_{\mu \nu}$ operators, while leaving the lower-dimension Higgs couplings in Eq.~\eqref{eq:1} unchanged relative to the SM prediction. The red-line shows the trajectory $\delta c_{\gamma} = 2/9 \delta c_{g}$ characteristic for top partners. The {\bf top right} plot characterizes composite Higgs models. The {\bf bottom}  plots characterize {\emph{top partner}} models where  only scalars and fermions with the same charge and color as the top quark contribute to the effective 5-dimensional operators, which implies the relation $\delta c_\gamma = (2/9) \delta c_g$.
In the bottom-right plot, the shaded region is 95\% CL excluded by monojet  searches at  the LHC.
 \label{fig:generalFit}}
\end{figure}

Since the ATLAS experiment did not provide the values of $R_{\gamma\gamma jj}$ for $m_h = 125 $ GeV, we estimate the best fit values and error bars from the p-values in Fig. 11 of \cite{Incandela:inc12}. Assuming the probability distribution function follows Gaussian statistics, the best fit rate, $\hat{\mu}$, and error bar, $\sigma$ can found by inverting the observed and expected  p-values ($p_0$):
\beq
1/\sigma = \Phi^{-1}(1-p_0^{\rm exp}), \qquad \hat{\mu}/\sigma = \Phi^{-1}(1-p_0^{\rm obs})
\label{eq:pvalue}
\eeq
where $\Phi(x)\equiv \frac{1}{2}(1-{\rm Erf}(x))$ is the cumulative distribution function. Assuming the significance across each channel adds in quadrature, the rates from the 10 diphoton channels can be related to the rates of the 9 inclusive channels and the dijet channel. This can be used to calculate $\hat{\mu}_{\gamma\gamma jj}$ and $\sigma_{\gamma\gamma jj}$,  after utilizing \eqref{eq:pvalue} to determine the other rates and error bars. For $m_h = 126.5 $ GeV, we find $\hat{\mu}_{\gamma\gamma jj} = 2.9 \pm 1.3$, whereas ATLAS reports  $\hat{\mu}_{\gamma\gamma jj} = 2.7 \pm 1.3$, which we take as validation of our prescription for determining the rates using the reported p-values.

With enough data from the LHC one could in principle perform a full seven-parameter fit, however for the time being we pursue a simpler approach.
Here we assume  $c_\tau = c_b$ and $c_{\rm inv}=0$, and study the LHC and Tevatron constraints on  the remaining parameter space. In this space, the best-fit points are
\beq
|\hat{c}_\gamma| =  1.2^{+0.7}_{-0.4},    \qquad  |c_V| = 1.2^{+0.4}_{-0.5}, \qquad  \left|{\hat{c}_g}/{ c_b} \right| = 0.8^{+0.7}_{-0.3}, \qquad |c_b| > 0.15.
\eeq
 Notice from \eqref{eq:4}-\eqref{eq:last} that $\chi^2$ mostly depends on the combination $\left|{\hat{c}_g}/{ c_b} \right| $, and thus does not strongly constrain $\hat{c}_g$ and ${ c_b}$ separately.  The corresponding error bars are 1$\sigma$, and have been derived assuming that the $\chi^2$ distribution around the minimum follows a $\Delta \chi^2$ distribution for 4 DOF.
Relative to the SM  point, $\Delta \chi^2 = \chi^2_{\rm SM} - \chi^2_{\rm min}   =  5.2$, which is consistent with the global minimum at $74\%$ CL.

We also  study the best-fit regions in new physics models where only two of the above parameters can be freely varied, while the remaining ones are fixed to the SM values. Sample results are displayed in Fig.~\ref{fig:generalFit}. The results are shown for 4 different sets of assumptions about the Higgs couplings that can be realized in concrete models. In each plot we show the $1\sigma$ constraints from the individual $R_{ii}$ while  the ``Combined'' region corresponds to $\Delta  \chi^2 < 5.99$, that is the 95\% CL favored region. An interesting feature of all these plots is the presence of two disconnected best-fit regions  \cite{Carmi:2012yp,Azatov:2012bz}.

The top left plot characterizes models in which loops containing beyond the SM fields contribute to the effective $h \, G_{\mu \nu}^a G_{\mu \nu}^a$ and $h \, A_{\mu \nu} A_{\mu \nu}$ operators, while leaving the lower-dimension Higgs couplings in Eq.~\eqref{eq:1} unchanged relative to the SM prediction.
Note that in this plot  the band corresponding to $R_{bb}^{\rm VH}$ is absent; that is done for aesthetic reasons since $R_{bb}^{\rm VH}$ is very weakly independent of $c_g$ and $c_\gamma$. The data prefer negative contributions to $c_g$ (decreasing the ggF production rate) and to  $c_\gamma$ (increasing the diphoton decay width).
An improvement of the fit with respect to the SM by $\Delta \chi^2 \sim 4$  is possible for large $\delta c_\gamma / \delta c_g$, which would require a loop contribution from a particle with a large charge-to-color ratio \cite{Bonne:2012im} ($Q_i^2 \simgt 2$ for the fundamental representation of $SU(3)_c$), or simply without color charge.
Another possibility is a particle with $\delta c_\gamma/\delta c_g \sim  0.25$ ($Q_i^2 \sim 1/2$ for the fundamental representation of $SU(3)_c$) giving a very large negative contribution to the effective gluon coupling,  tuned such that $\delta c_g  \sim -2 c_{g,\rm SM}$.
Finally, a number of  particles with different electric and/or color charges could be involved in such a way that their net contribution to $\delta c_g$ (but not to $\delta c_\gamma$) approximately cancels.

In the remaining 3 plots we fix $\delta c_\gamma = (2/9) \delta c_g$, which is the case in  {{top partner}} models where  only scalars and fermions with the same charge and color as the top quark contribute to these effective five-dimensional operators.
In the  top-right plot, the couplings to all the SM fermions, including the one to  the top quark in the UV completed model is assumed to be rescaled by $c_b$, producing the corresponding shift of  $c_g$ and $c_\gamma$ in our effective theory. 
Moreover, the coupling to $W$ and $Z$ is independently rescaled by $c_V$.
This is  inspired by  the composite Higgs scenario \cite{Giudice:2007fh}, in which context  a part of the parameter space with $c_b < 1$ and $c_V < 1$ can be reached in specific models.
The presence of two disconnected best-fit regions  reflects the degeneracy of the relevant Higgs rates in the $VV^*$ and $b \bar b$ channels under the reflection $c_b \to -c_b$,  which is broken only by $R_{\gamma \gamma}$. 
A good fit  is possible in the $c_b < 0$ region, although it may be difficult to construct a microscopic model where such a possibility is realized naturally.
It is worth noting that  the fermiophobic Higgs scenario, corresponding to $c_b = 0$ and  $c_V =1$, is  disfavored by the data (more generally, the fermiophobic line $c_b = 0$ is disfavored for any $c_V$).

The bottom-left plot  demonstrates  that the current data show a preference for a slightly enhanced Higgs coupling to the electroweak gauge bosons, $c_V> 1$.
Several well-studied models such as the MSSM or the minimal composite Higgs (and more generally, models with only SU(2) singlets and doublets in the Higgs sector), predict $c_V \leq 1$. If $c_V > 1$ is confirmed by more data, it would point to a very specific and interesting direction for electroweak symmetry breaking \cite{Falkowski:2012vh}.

In the bottom-right plot we relax one of our assumptions and allow for invisible Higgs decays.
Opening the invisible channel reduces the visible rates, therefore it is in tension with the observations in the diphoton, $VV^*$, and $b \bar b$ channels.
Therefore,  the data disfavor an invisibly decaying Higgs with ${\rm Br}_{inv} > 20$\% unless there is a significantly negative $\delta c_g$ (and, in the present case, the correlated negative contribution to $\delta c_\gamma$ enhancing the diphoton Higgs rate) in which case  ${\rm Br}_{inv}$ as big as $65$\% is allowed.
Note that invisible Higgs production is directly constrained by  monojet  searches at the LHC \cite{Djouadi:2012zc}.
As a result, interesting regions of the ${\rm Br}_{inv}$-$\delta c_g$ parameter space where the Higgs production cross section  is  enhanced compared to the SM  are already  excluded at  95\% CL.

\section{Simplified Models}
\label{sec:simplified}

In this section we discuss simple models where the interactions  of the Higgs boson with matter may deviate from the SM predictions.
In each case, we first map these models to our effective Lagrangian in \eqref{eq:1}, which facilitates an extraction of the observable rates.
The models we consider introduce the minimal number of new degrees of freedom.
We pay special attention to whether the new degrees of freedom allow for an enhanced diphoton rate, which is hinted by the data.   We also study whether the models may improve the Higgs naturalness, i.e., whether they can cancel the quadratic divergent contributions to the Higgs mass induced by the SM particles.

 The study of such simplified models allows one to identify the required Weak-scale physics which can better account for the Higgs data.   More complete models, which are typically constrained by additional experimental data can then be derived, and we postpone the detailed study of complete models to an upcoming publication.

\subsection{Single Partner Models}

As a first exercise, consider a class of simplified models with only one new degree of freedom coupled only to the Higgs boson.
The new degree of freedom, below referred to as the partner,  could be a  scalar $S$, a Dirac fermion $f$, or a vector boson $\rho$, carrying charge and/or color, and coupled to the Higgs as in Eq.~\eqref{eq:partners},
\beq
\cl = -  c_s {2 m_s^2 \over v}  h  S^\dagger S  - c_f {m_f \over v} h \bar f f  +  c_\rho {2 m_\rho^2 \over v} h \rho_\mu^\dagger  \rho_\mu  \,.
\eeq
Here appropriate index contractions are implicit for colored partners. For $c_i = 1$ the mass of the partner originates completely from electroweak symmetry breaking with a single Higgs, but we do not require this to be the case in general.
For simplicity we assume in this subsection   that the partner does not mix with the SM fields.   This can be arranged, for example,  by imposing a conserved $Z_2$ symmetry.  We relax this assumption in the subsections below.

\begin{figure}
\bc
\includegraphics[width=0.42\textwidth]{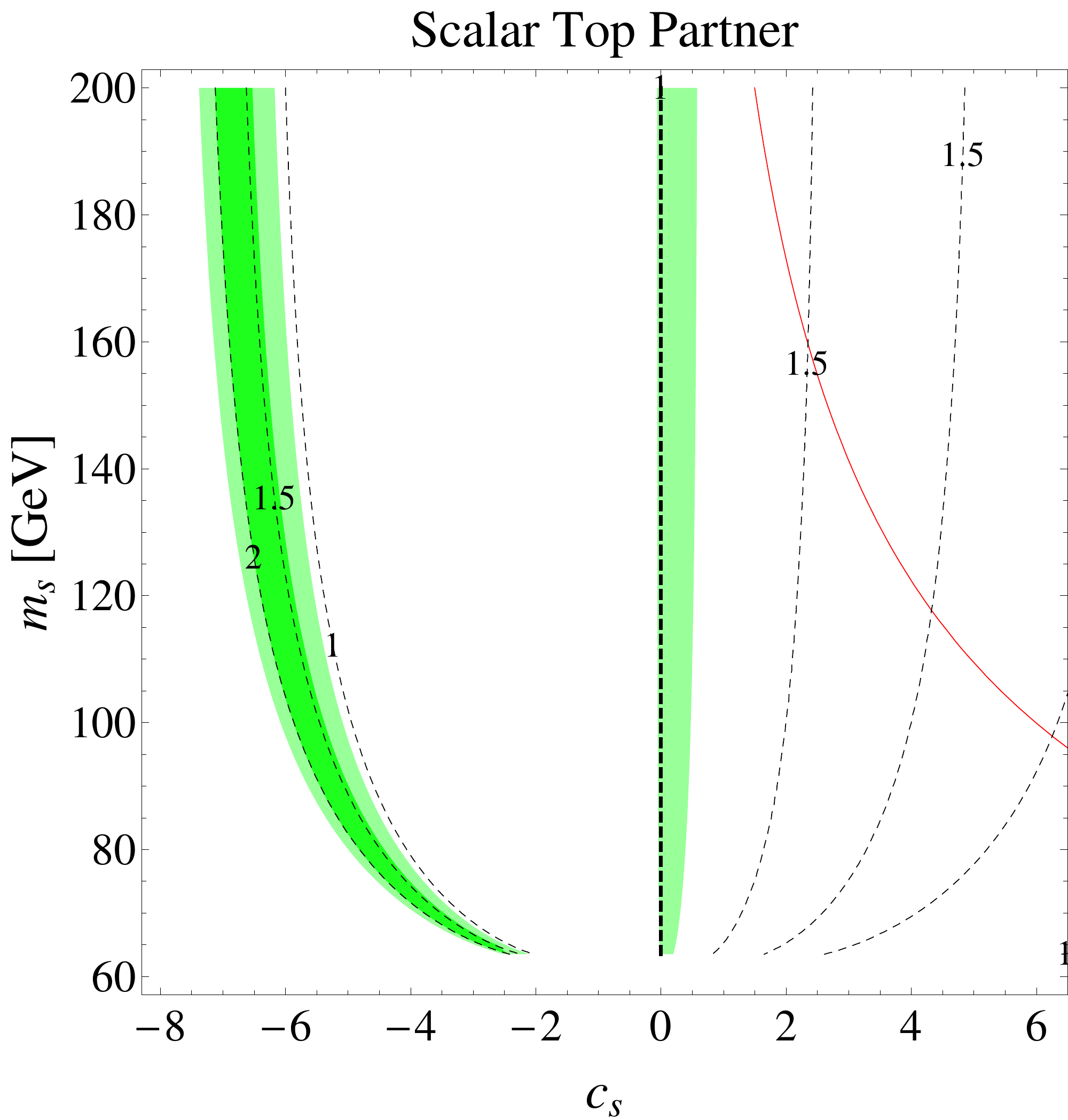}\quad \includegraphics[width=0.42\textwidth]{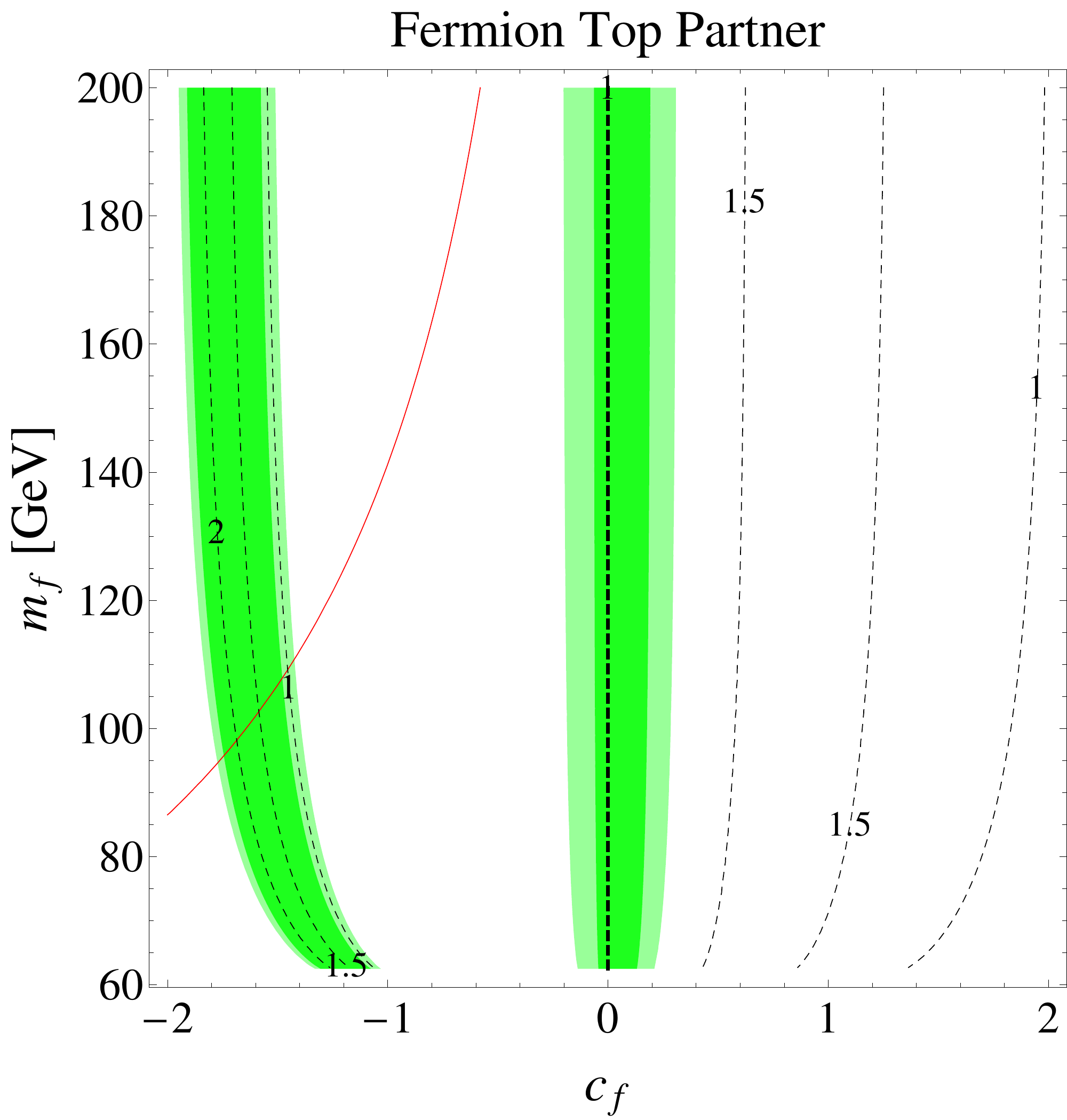}\\
\quad
\includegraphics[width=0.42\textwidth]{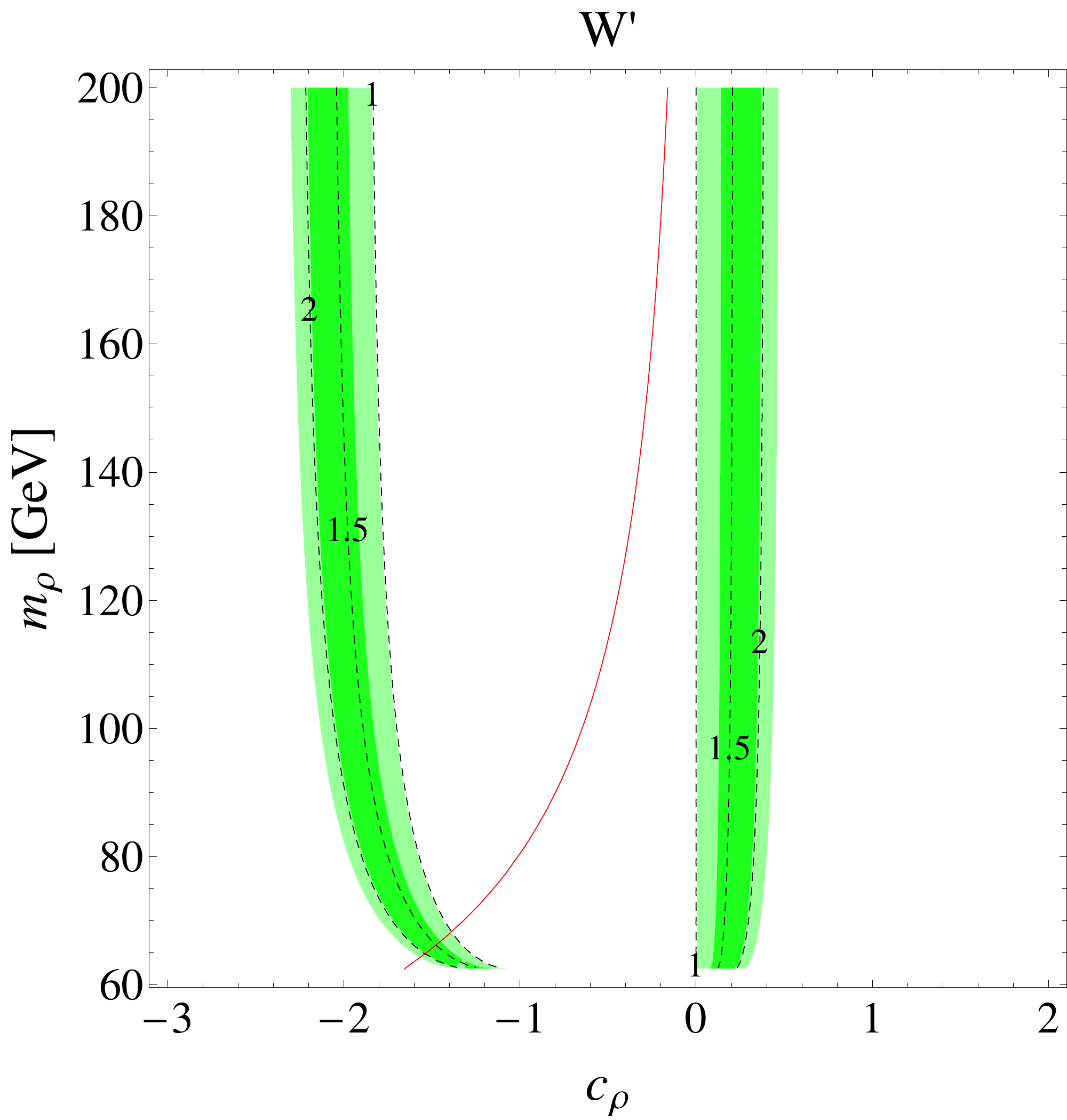} \quad \includegraphics[width=0.41\textwidth]{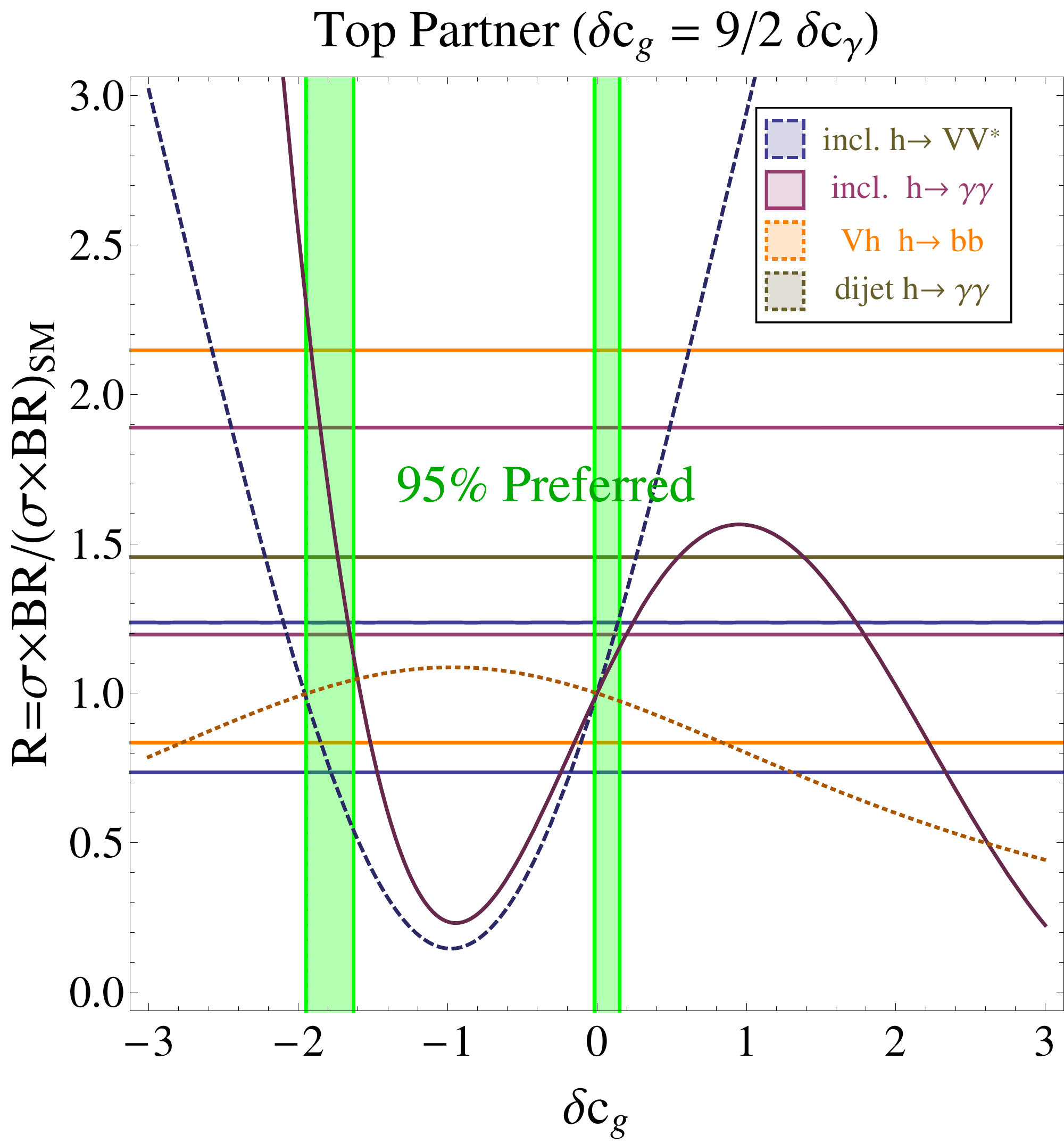}
\ec
\caption{\small
 Best fit regions in the $c_i$-$m_i$ plane, assuming $m_h = 125$ GeV for the scalar top partner ({\bf top-left}), fermionic top partner ({\bf top-right}) and vector $W$-partner ({\bf bottom-left}).
Shown are $68\%$ (darker green) and $95\%$ CL (lighter green) regions. The dashed curves are for constant $R_{\gamma\gamma}^{\rm incl}$, Eq. \eqref{eq.Rgg}, while the red curve is where a single partner is improving the naturalness of the SM, Eqs. \eqref{eq.cs.natural}, \eqref{eq.cf.natural}, \eqref{eq.crho.natural}.
The {\bf bottom-right} image shows the constraints for $m_h = 125$ GeV, for top partner models, i.e. $\delta c_{\gamma} = 2/9 \delta c_{g}$. The three bands show the 1$\sigma$ allowed regions
for $R_{bb}^{VH}$, $R_{\gamma\gamma}^{\rm incl.}$, and $R_{ZZ}^{\rm incl.}$ channels. The three curves show the theoretical predictions as a function of $\delta c_{g}$ for each channel. Only 3 channels are shown, but all channels are included. The
green shaded region shows the 95\% CL experimental preferred region.
 \label{f.toppartner}}
\end{figure}

Integrating out the partner affects the dimension-5 Higgs couplings to gluons and photons $c_g$ and $c_\gamma$, while keeping the remaining parameter in \eqref{eq:1} at the SM values $c_V=c_b=c_c=c_\tau =1$.
 The ratio $c_g/c_\gamma$ is determined by the electric charge and the color representation of the partner.
As an illustration we present our results for the following three  cases:
 \bi
\item {\bf Scalar top partner.} Color triplet, charge 2/3 scalar, contributing as
\begin{align}
\delta c_g   &=  {c_s \over 4} A_s(m_h^2/4 m_s^2)\,,
\\
\delta c_\gamma   &=  {1 \over 18} c_s A_s(m_h^2/4 m_s^2)\,.
\end{align}
The partner exactly cancels the quadratic divergence from the top quark for
\beq
c_s =  \frac{2 m_t^2}{m_s^2}\,. \label{eq.cs.natural}
\eeq
\item {\bf Fermionic top partner.} Color triplet, charge 2/3 fermion, contributing as
\begin{align}
\delta c_g  & =  c_f A_f(m_h^2/4 m_f^2)\,,
\\
\delta c_\gamma  &=  {2 \over 9}c_f A_f(m_h^2/4 m_f^2)\,.
\end{align}
The partner exactly cancels the quadratic divergence from the top quark for
\beq
c_f = -{2 m_t^2 \over m_f \left(m_f + \sqrt{2 m_t^2 + m_f^2} \right )}\,,  \label{eq.cf.natural}
\eeq
and, in that case, in the limit $m_f \gg m_t$ one has $\delta c_g = 9 \delta c_\gamma/2 \approx -m_t^2/m_f^2$.

\item {\bf W prime.}  Color singlet, positively charged massive vector, contributing as
\begin{align}
\delta c_g   &= 0\,,
\\
\delta c_\gamma    &=  -{7 \over 8} c_\rho A_v(m_h^2/4 m_\rho^2)\,.
\end{align}
The partner exactly cancels the quadratic divergence from the W boson for
\beq
c_\rho = \frac{m_W^2}{m_\rho^2}\,.  \label{eq.crho.natural}
\eeq
\ei

The best fit  68\% CL and 95\% CL regions for these 3 examples are shown in \fref{toppartner}.
In each case, the model is defined by 2 parameters: the coupling $c_i$ and the mass $m_i$, but only one combination influences the effective theory parameters.
Consequently, the best fit region  in the $c_i$-$m_i$ plane corresponds to  a line of minimum $\Delta \chi^2$.
For a scalar, the best fit occurs for a large negative coupling $c_s$, which is incompatible with naturalness in this simple set-up.
To improve the fit to the Higgs data with scalar partners one needs at least two of them (as in supersymmetric theories) and with a large mixing, so that the sign of coupling of the lighter partner to the Higgs is flipped due to the mixing angle.
Note that in that case the lighter scalar actually worsens the quadratic divergence of the SM, and the restoration of naturalness is postponed to the higher scale where the heavier scalar intervenes.

For the fermionic top partner  case the situation is different as  a negative coupling is consistent  with naturalness.
Therefore a single fermionic  top partner may improve the fit to the Higgs data and the naturalness at the same time provided the partner is light, with mass in the narrow range of 95 GeV $< m_f <$ 115 GeV.   Such low masses are likely to be excluded by direct searches and precision measurements in  specific more complete models.
In the vector $W$ partner case, there are 2 separate best fit regions with the same $\chi^2_{\rm min}$: one where $c_\rho$ is relatively small and positive, where the partner interferes constructively with the SM $W$ boson, and the other where $c_\rho$ is large and negative, so that the partner "overshoots" the SM $W$ contribution.
Only the latter region can overlap with the curve where the partner exactly cancels the quadratic divergence from the $W$ boson.

To summarize, for all models an enhanced diphoton decay rate can be obtained.
In the scalar and fermionic case  one needs a sizable negative coupling $c_i$, while in the vector case a moderate positive coupling suffices.
The diphoton rate is indicated by  the constant contours shown in Fig.~\ref{f.toppartner}. For the top partner case, the fit to data is  improved relative to the SM  as  $\Delta \chi^2 = \chi^2_{\rm min}-\chi^2_{\rm SM} = 3.8$ for only 1 {\it{dof}}.

\subsection{Composite Higgs Models}
\label{sec:composite}

\begin{figure}
\includegraphics[width=0.45\textwidth]{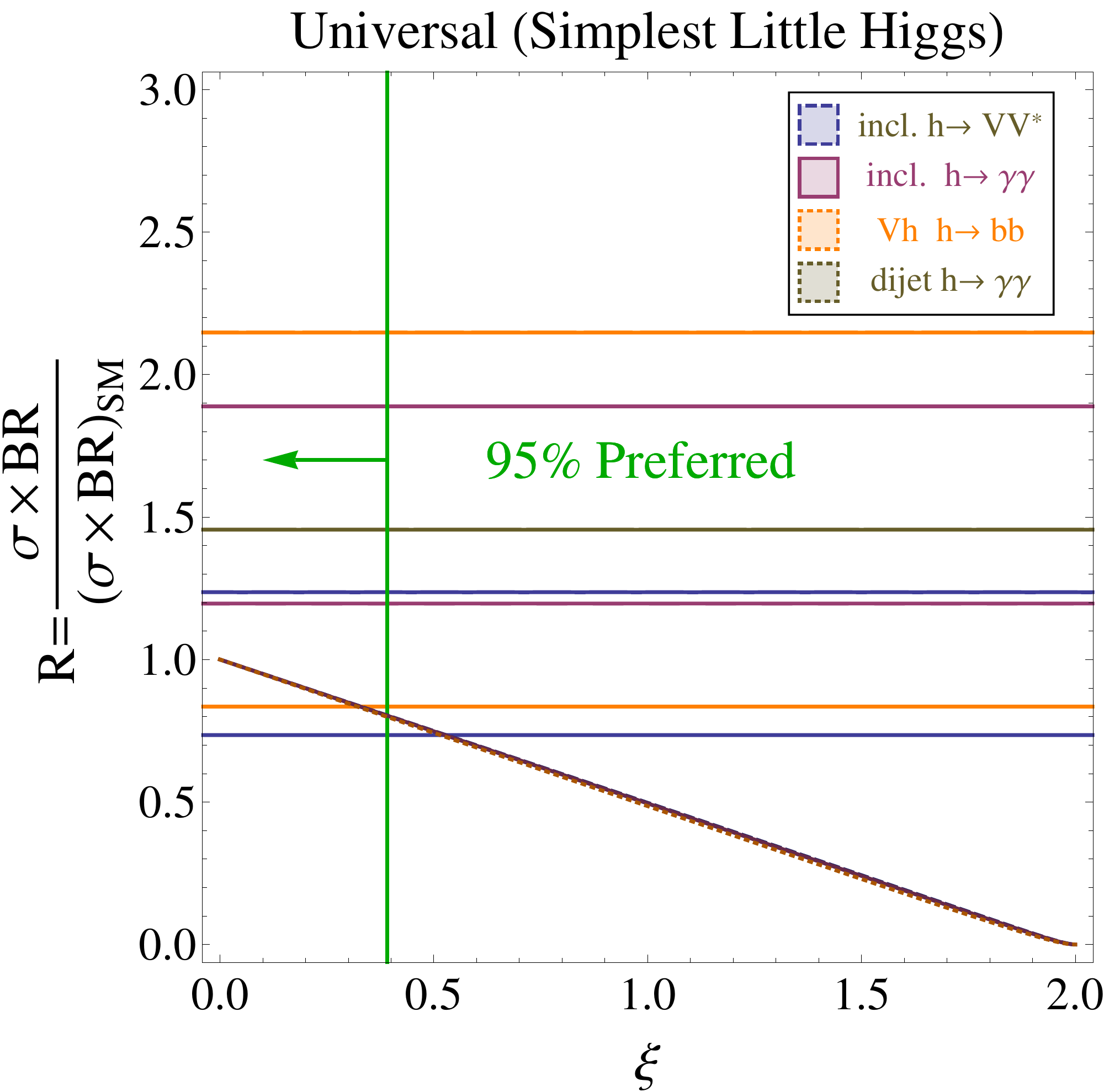}
\quad
\includegraphics[width=0.45\textwidth]{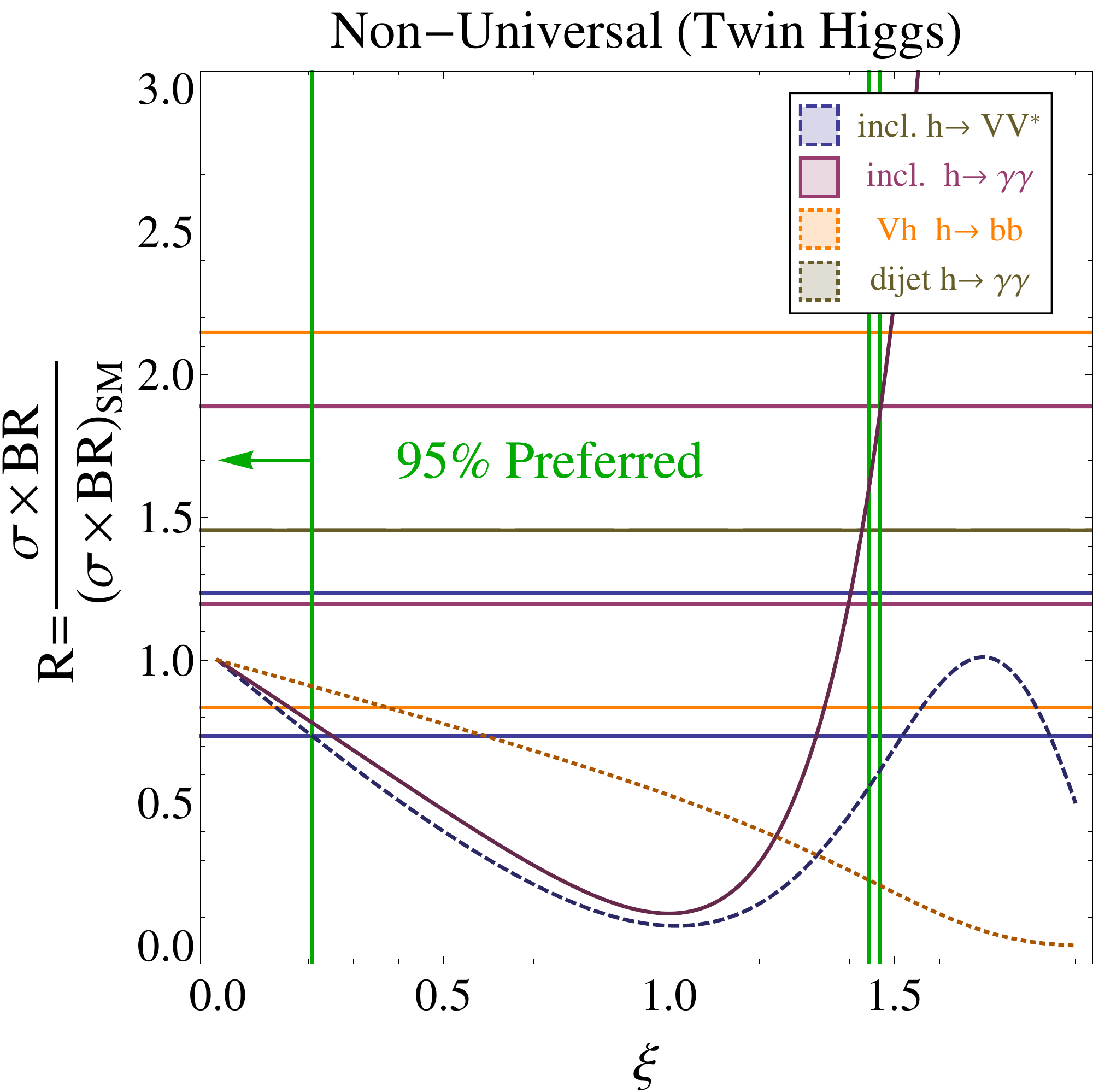}
\caption{\small
Constraints for the Simplest Higgs model ({\bf left}) and the Twin Higgs model ({\bf right}) assuming $m_h = 125$ GeV. The three bands show the 1$\sigma$ allowed regions
for $R_{bb}^{VH}$,  $R_{\gamma\gamma}^{\rm incl.}$,  and $R_{ZZ}^{\rm incl.}$ channels. The three curves show the theoretical predictions as a function of $\xi$ for each channel. Only 3 channels are shown, but all channels are included. The
green vertical lines show the 95\% CL experimental preferred region.
 \label{f.littlehiggs}}
\end{figure}

The single fermion partner model described in the previous subsection is a special case of models with a more general set of couplings of fermions to the Higgs.   In particular, consider a vector-like pair of top partners, $T$ and $T^c$, which interact with a pseudo-Goldstone Higgs.   There are several possibilities for the interactions, depending on the type of model under considerations.  Here we consider two possibilities:
\beq
\label{eq:topSLH}
-\cl_{top} = y f \sin(|H|/f) t  X^c  +  y f \cos(|H|/f) T  t^c  + M' T T^c +\hc  \,.
\eeq
where either $X^c = t^c$ or $X^c = T^c$.   The former occurs, e.g., in the Simplest Little Higgs model with an $[SU(3)/SU(2)]^2$ coset~\cite{Schmaltz:2004de} or in the minimal composite Higgs with $SO(5)/SO(4)$ coset structure~\cite{Agashe:2004rs}.   The latter case, is encountered in the left-right Twin Higgs model~\cite{Chacko:2005pe}.

For sufficiently large $M^\prime$, the heavy partners can be integrated out and the low energy couplings to gluons and photons are found to be,
\begin{eqnarray}
{c_g \over c_{g,\rm SM}}  =  {c_\gamma \over c_{\gamma,\rm SM}}
=\left\{ \begin{array}{lll}
\sqrt{1 - {\xi/ 2} }&\qquad& X^c = t^c\\
{1  - \xi \over  \sqrt{1 - \xi / 2 } } &\qquad& X^c = T^c
\end{array}\right. ,
\end{eqnarray}
where $\xi = v^2/f^2$.  Similarly, for the models above one finds,
\beq
c_V = c_b = \sqrt{1 - {\xi/2} },
\eeq
Thus we see that the couplings of the pseudo-Goldstone Higgs may exhibit a universal or non-universal suppression which depends on the single parameter, $\xi$, and are independent on the specific details within the top sector (masses and mixing, for example).

In Fig.~\ref{f.littlehiggs}, we show the production and decay rates of the above models as a function of $\xi$.   We find that for a 125 GeV Higgs boson, $\xi$ is constrained at 95\% CL to be $\xi<0.4$ in the universal suppression case,  and $\xi<0.2$ in the non-universal case.

\subsection{Dilaton}

\begin{figure}
\includegraphics[width=0.45\textwidth]{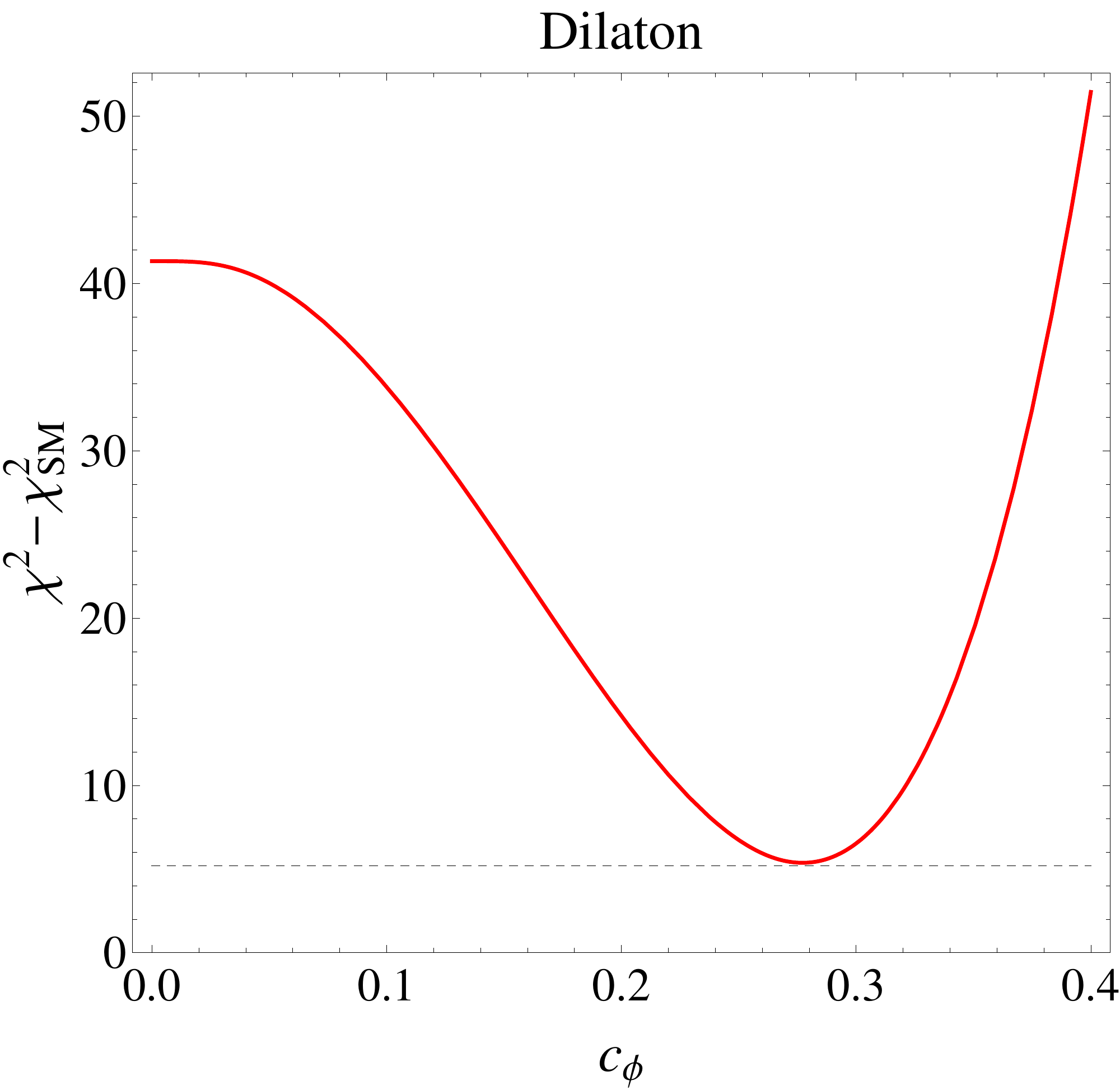}
\quad
\includegraphics[width=0.4\textwidth]{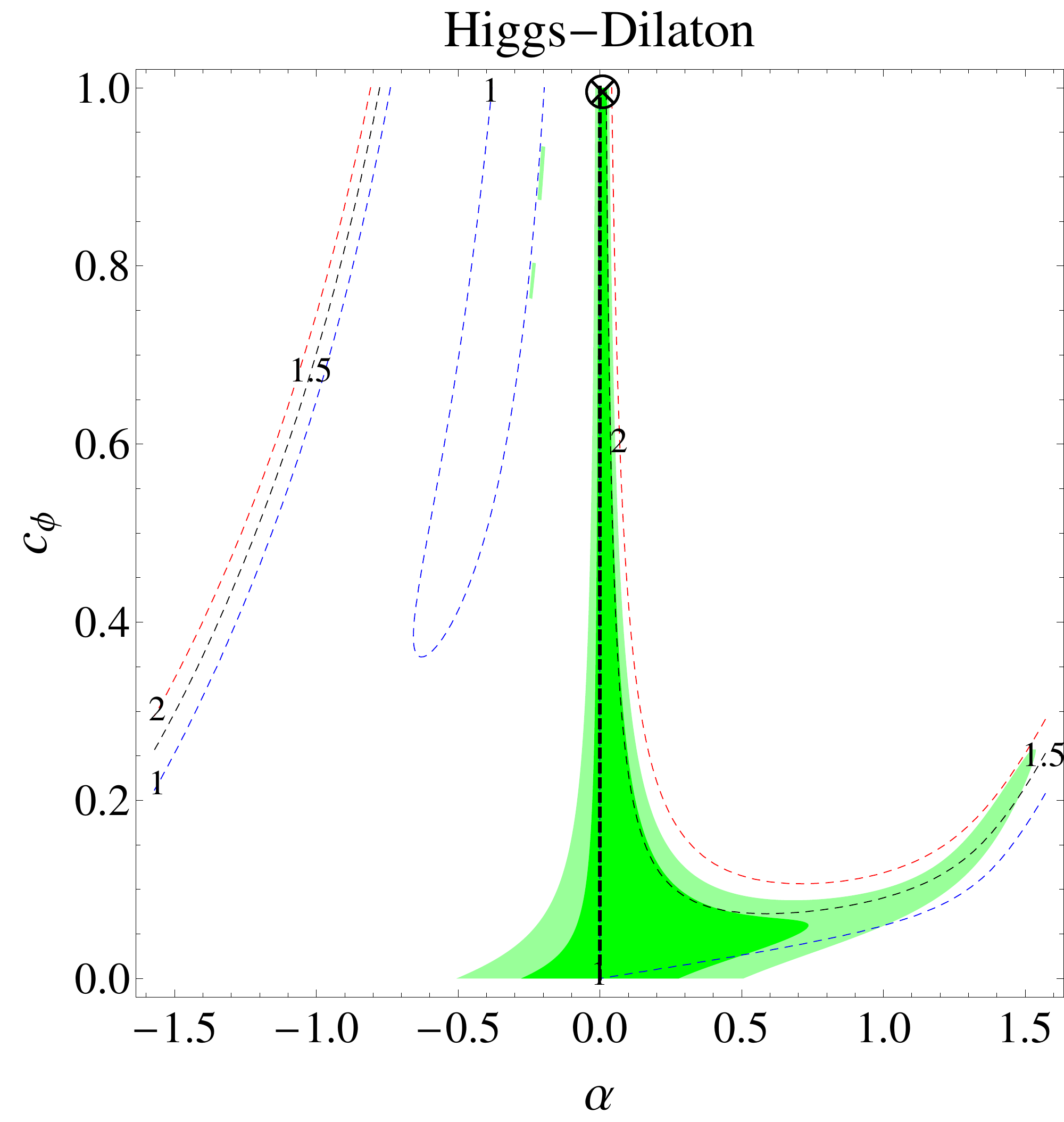}
\caption{\small
\textbf{Left:} The difference between the $\chi^2$ of the dilaton model and the $\chi^2$ of the SM as a function of the parameter $c_\phi$.
At the best fit point around $c_\phi \simeq 0.27$, the $\chi^2$ of the dilaton model is larger by 5.2 units compared with the SM, indicating that the dilaton model always fits the data worse than the SM.
\textbf{Right:} The favored region at 68\% CL (Darker) and 95\% CL (Lighter) for the $125$ GeV resonance being a mixture of the SM Higgs and a dilaton.
The best fit occurs along the line  $\alpha =0$ corresponding to a pure SM Higgs.
The dashed lines show contours of constant $R_{\gamma \gamma}$.
 \label{f.dilaton}}
\end{figure}

Here we study the hypothesis that the 125 GeV resonance discovered  at the LHC  is a  {\em dilaton} \cite{Goldberger:2007zk,Giardino:2012ww},  that is to say a (pseudo)-Goldstone boson of the spontaneously broken conformal symmetry.  That implies that it couples to the trace of the energy-momentum tensor of the SM matter,
\beq
\cl = c_\phi {\phi \over v} \left ( 2 m_W^2  W_\mu^+ W_\mu^-  +   m_Z^2  Z_\mu Z_\mu - \sum_{f \in SM} m_f \bar f f  \right ) \sim \phi T_\mu^\mu
\eeq
In other words, the lower-dimension couplings are the same as for the SM Higgs bosons, up to an arbitrary overall rescaling factor $c_\phi$.
Mapping to the effective Lagrangian \eqref{eq:1}, we have,
\beq
c_V = c_b = c_\tau = c_c = c_\phi\,.
\eeq
The dimension-5 coupling do not however follow the overall rescaling because they are affected by integrating out new heavy degrees of freedom needed to restore conformal invariance at high energy.
Using conformal symmetry,  these UV contributions to $c_g$ and $c_\gamma$ can be related to  the QCD and electromagnetic beta functions:
$c_g = - c_\phi 3 b_0^{\rm QCD}/2$, $c_\gamma = - c_\phi b_0^{\rm EM}/8$.
Since above the scale of the top mass one has $b_0^{\rm QCD} = -7$ and $b_0^{\rm EM} = 11/3$, integrating out the top quark one finds
\beq
\label{e.dcgDilaton}
\delta c_g =  {21 \over 2} c_\phi +  \left(c_\phi - 1 \right ) A_f(\tau_t) \, ,
\qquad
\delta c_\gamma = - {11 \over 24} c_\phi +  \left(c_\phi - 1 \right ) A_f(\tau_t) \, .
\eeq
Thus, for $c_\phi > 0$, the effect of the dilaton is to increase the $H\to \gamma \gamma$ width, which is favored by the data, but at the price of increasing the ggF production rate, which is  disfavored.
We therefore find that at the best fit point at $c_\phi \simeq 0.27$, the $\chi^2$ of the dilaton model is larger by 5.2 units than $\chi^2_{SM}$, as can be seen in the left panel of \fref{dilaton}.
Thus, the simple dilaton interpretation of the 125 GeV resonance is not favored by  the data.
Allowing for negative $c_\phi$ does not change that conclusion, as the rates are symmetric under $c_\phi \to - c_\phi $.

More generally, one may consider the dilaton mixed with the SM Higgs boson, with the mixing angle denoted by $\alpha$.
In that case the effective theory parameters are given by
\bea
\label{e.dcgHiggsDilaton}
\delta c_g &=&  {21 \over 2} c_\phi \sin \alpha +  \left(c_\phi \sin \alpha  + \cos \alpha  - 1 \right ) A_f(\tau_t) \, ,
\nn
\delta c_\gamma &=&  - {11 \over 24} c_\phi \sin \alpha +  \left(c_\phi \sin \alpha  + \cos \alpha  - 1 \right ) A_f(\tau_t) \, .
\eea
The best fit regions in the $\alpha$--$c_\phi$ plane are shown in the right panel of \fref{dilaton}.
Again, the best fit is obtained along the SM line $\alpha = 0$.

We conclude, therefore, that at the moment, there are no hints of a dilaton nature in the 125 GeV resonance, although a large Higgs-dilaton mixing angle cannot be excluded at present.

\section{Extended Higgs Sectors}
\label{sec:extended}
We next turn our attention to extended Higgs sectors with one or more additional scalars beyond the $125$ GeV Higgs.   As we show below, several scenarios are possible, allowing for an enhanced diphoton or ZZ and WW rate.

\subsection{The Two-Higgs Doublet Model}

\begin{figure}[tb]
\bc
\includegraphics[width=0.45\textwidth]{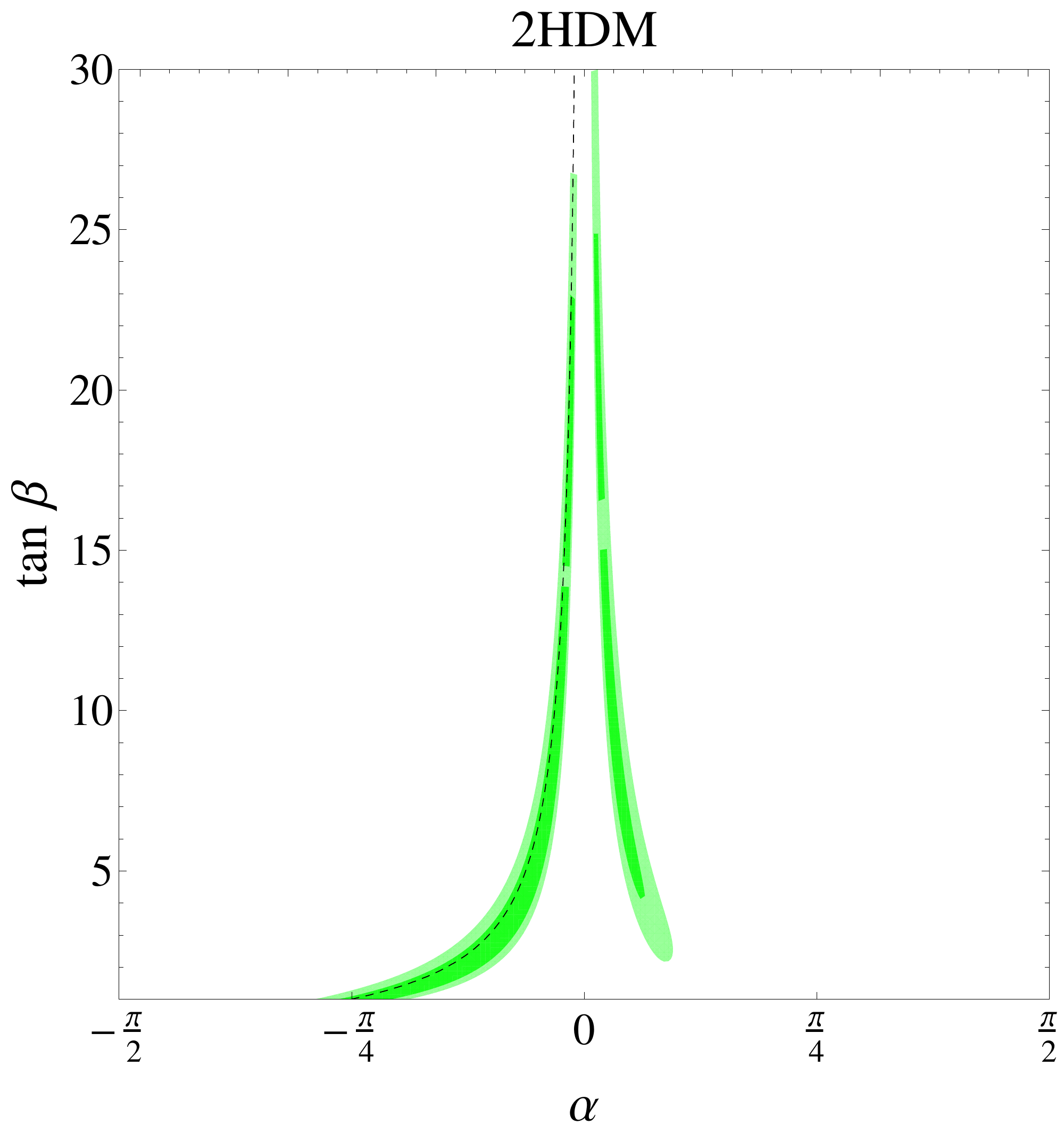}
\ec
\vspace{-1cm}
\caption{\small Favored region at 95\% CL, of the 2HDM in the $\alpha-\tan \beta$ plane.
For $\sin \alpha < 0$ the favored region  is  concentrated around the decoupling limit (dashed line), $\alpha = \beta - \pi/2$ where all couplings are SM-like.  Conversely, for $\sin \alpha >0$ the favored region lies around  $\alpha = \beta + \pi/2$ where the top Yukawa coupling is SM-like, but $c_V = -1$.
 \label{2HDM}}
\end{figure}

Consider the Type II 2-Higgs Doublet Model (2HDM), with 2 Higgs doublet fields $H_u$, $H_d$:
\bea
H_u= \left (H_u^+, {v_u + H_u^0 \over \sqrt 2 } \right ) \,, \qquad\ H_d= \left( {v_d + H_d^0 \over \sqrt 2},H_d^- \right )\,.
\eea
The doublet $H_u$  couples to up-type quarks, and $H_d$ to down-type quarks and leptons. The  ratio of the two VEVs
is $\tan \beta \equiv {v_u}/{v_d}$, and by convention $0 < \beta < \pi/2$.
The CP-even mass eigenstates are mixtures of the neutral components
\begin{eqnarray}
\begin{pmatrix}
H\\
h\\
\end{pmatrix}     =
\begin{pmatrix}
\cos \alpha & \sin \alpha\\
-\sin \alpha & \cos \alpha\\
\end{pmatrix}
\begin{pmatrix}
{\rm Re} (H_d^0)\\
{\rm Re}(H_u^0)\\
\end{pmatrix},
\end{eqnarray}
with  $\alpha$ the mixing angle.
We identify $h$ as the 125 GeV Higgs.
The tree level Higgs couplings to the fermions and vectors are given by
\beq
c_V  = \sin(\beta - \alpha)\,,
\quad
c_b = -{\sin \alpha \over \cos \beta}\,,
\quad
c_g = 9/2 c_{\gamma} \simeq  {\cos \alpha \over \sin \beta}\,,
\eeq
where the effect on $c_g$ arises because of the modified Higgs coupling to the top quark.

In Fig.~\ref{2HDM} we show the constraints in the $\alpha$-$\tan \beta$ plane.
In accordance with current direct bounds~ \cite{Deschamps:2009rh}, we assume that the charged Higgs is heavy enough so as to contribute negligibly to $c_{\gamma}$, and $\tan \beta \gtrsim 0.3 $, so that the top Yukawa coupling does not run to a Landau pole at $\mu \sim$ TeV.
The best fit approximately corresponds to the decoupling limit  $\alpha = \beta - \pi/2$ ($\tan \alpha = - \cot \beta$) where all couplings are SM-like, in particular $c_V\approx1$.  The minimum $\chi^2$ is  roughly the same as in the SM,
$\Delta \chi^2 = \chi^2_{\rm min} - \chi^2_{\rm SM} \simeq 0$.  Another favored region is for $\alpha > 0$, where $c_V$ is still close to $1$ and the sign of $c_b$ is flipped.

\subsection{Simplified MSSM}
\begin{figure}[tb]
\vspace{0cm}
\bc
\includegraphics[width=.45\textwidth]{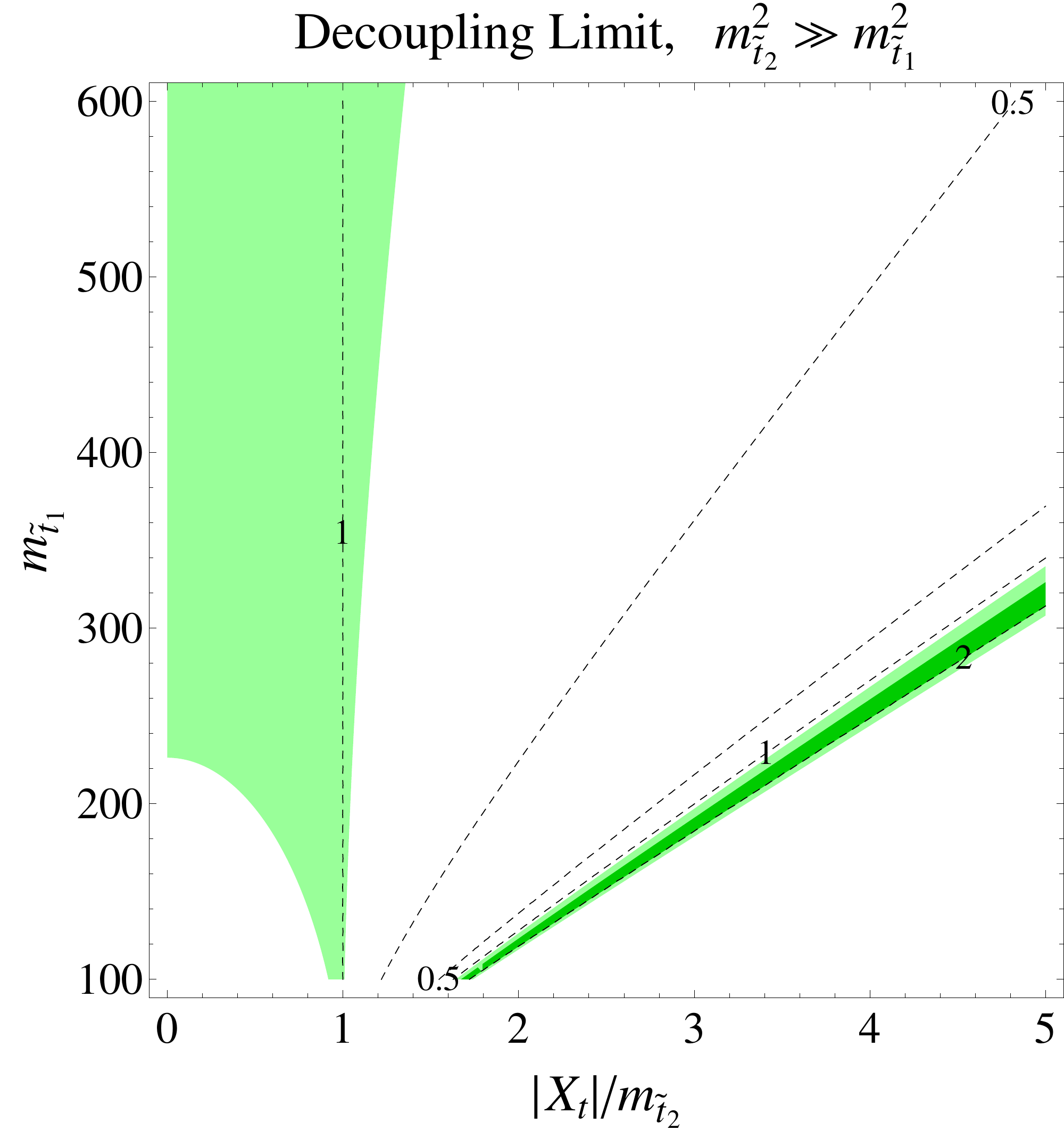}
\quad
\includegraphics[width=.45\textwidth]{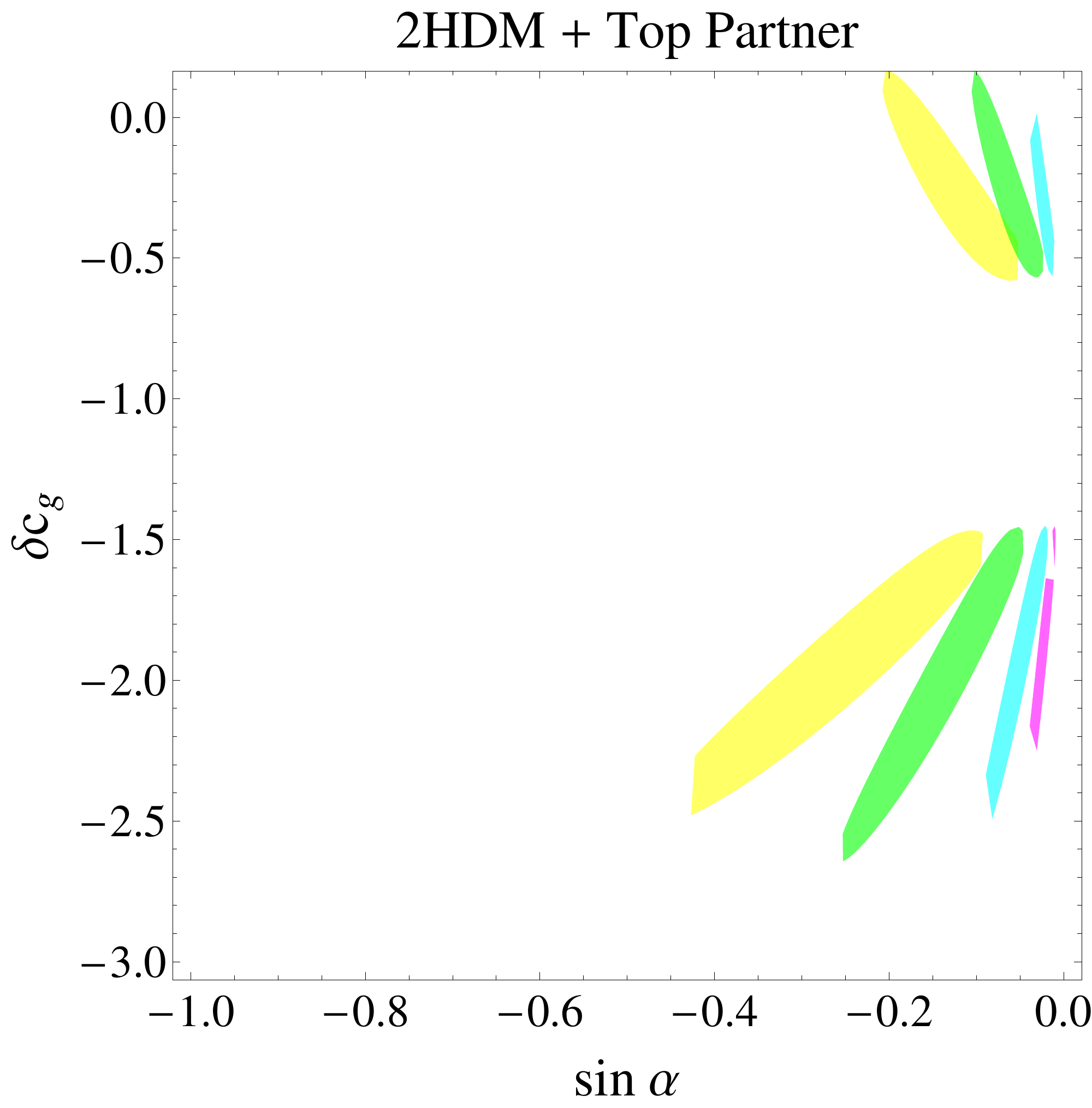}
\ec
\vspace{-.5cm}
\caption{\small\textbf{Left:} The favored region at 68\% CL (Darker) and 95\% CL (Lighter) for $m_h = 125$ GeV in the two scalar model,  with $m_{{\tilde{t}_2}} \gg m_{{\tilde{t}_1}}$, and assuming the decoupling limit of the 2HDM. The dashed lines show contours of constant $R_{\gamma \gamma}$.
\textbf{Right:} Allowed regions at 95\% CL obtained for the 2HDM + two stops model. The different colors correspond to different values of $\tan \beta$. The regions are: $\tan \beta=5$ (yellow), $\tan \beta=10$ (green), $\tan \beta=25$ (blue) and $\tan \beta=50$ (purple). The best fit points corresponds approximately to the decoupling limit.
\label{fig:SUSY}
}
\end{figure}

The next example we consider is a simplified model of the MSSM: two Higgs doublet plus 2 stops defined as scalars with the same color and charge as the top quark.
The Higgs doublets are defined as in the previous section, but now $\alpha \in (-\pi/2,0)$.
Consider the stops $\ti t$,$\ti t^c$ with the mass terms of the form
\beq
-\cl_{stop} =   |\ti t|^2 \left ( \ti m^2  + y_t^2 |H_u|^2 \right )  +   |\ti t^c|^2 \left ( \ti m_c^2  + y_t^2 |H_u|^2 \right )
+  y_t |H_u| X_t \left ( \ti t \ti t^c + \hc \right),
\eeq
where $y$ is the top Yukawa coupling.
This is equivalent to the stop sector of the MSSM neglecting the  (sub-leading)  D-terms contribution to the stop masses.
The left-handed and right-handed stops mix in the presence of $X_t$, which in the MSSM  is given by  $X_t = |A_t - \mu \cot\beta|$.

We begin by considering the decoupling limit, $m_A \gg m_h$.  In that case, the change in rates are controlled by the stop spectrum.
Denoting the two mass eigenvalues by $m_{\tilde{t}_{i}}$, and the left-right mixing angle by $\theta_t$, one has
\beq
m_t X_t = \frac{1}{2}\left( m_{\tilde{t}_2}^2 - m_{\tilde{t}_1}^2 \right) \sin 2 \theta_t,
\eeq
where, by convention,  $m_{\tilde{t}_1} \le m_{\tilde{t}_2}$.
For $m_{\tilde{t}_{i}} \gg m_h/2$, integrating out the stops shifts the effective dimension-5 operators as
\beq
\label{e.SUSYc}
{c_g \over c_{g,\rm SM}}  =  {c_\gamma \over c_{\gamma,\rm SM}}    =
1 + {1 \over 4} \left({m_t^2 \over  m_{\tilde{t}_1}^2}   + {m_t^2
\over m_{\tilde{t}_2}^2} - {m_t^2 X_t^2\over  m_{\tilde{t}_1}^2 m_{\tilde{t}_2}^2} \right)\,.
\eeq
For zero mixing, the stops always  interfere constructively with the top contribution (and destructively with the $W$-contribution) to $\hat{c}_\gamma$. Assuming that both stops are heavier then about 100 GeV,  we see from Eq.~\eqref{e.SUSYc} that the contribution to  $c_g$ is bounded above by $\delta c_g \lesssim 1.5$. For large $X_t$, the sign of the contribution from stops can flip, and a significant enhancement of the diphoton width is possible.
Since $X_t$ needs to satisfy the bound $2 m_t |X_t| <( m_{\tilde{t}_2}^2 - m_{\tilde{t}_1}^2 )$, it can be large provided the two eigenstates are split.

In Fig.~\ref{fig:SUSY}  we illustrate the impact of the LHC Higgs data on the parameter space of the simplified MSSM model.
The left  plot shows the preferred region in the $m_{\tilde{t}_1}$--$X_t / m_{\tilde{t}_2} $ plane, assuming the decoupling limit and $m_{\tilde{t}_2} \gg m_{\tilde{t}_1}$ so that the heavier stop eigenstate does not contribute to the effective operators.
In this case, the Higgs data only constrains one combination of parameters, that is $c_g$ in Eq.~\eqref{e.SUSYc}. As can be seen in Fig.~\ref{fig:SUSY} , the data strongly prefers large mixing, where the mixing dominates the stop contribution to $c_g$ and $c_\gamma$. The best fit contour corresponds to $\delta c_g = -1.78$ and $\Delta \chi^2 = \chi^2_{\rm min} - \chi^2_{\rm SM} = 3.8$ for 1 DOF, thus the data strongly favors the top partner scenario.
On the other hand, the no mixing scenario,  $X_t =0$, is strongly disfavored, since $\delta c_g \ge0$.  Indeed in that case, the stop contribution can only enhance the gluon fusion cross section while decreasing the diphoton branching ratio, which does not fit well the LHC data.

Going back to the more general situation where the Higgs sector is away from the decoupling limit,
 the right plot of Fig.~\ref{fig:SUSY} shows the constraints from the Higgs data in the $\delta c_g - \sin\alpha$ plane for different values of $\tan\beta$. Although the parametrization is motivated by the MSSM with light stops, it is also applicable to any 2HDM model with  top partners. There are 3 independent parameters which are constrained by the Higgs measurements: $\tan\beta$, $\sin\alpha$, and $\delta c_g $. Even with the additional degrees of freedom, the Higgs fit to the data is not improved relative to taking the decoupling limit. This is simply a consequence of the fact that the data strongly points to $c_V \sim 1$, which can only be achieved in the decoupling limit.

\subsection{The Doublet-Singlet Model (Enhanced $c_{\gamma}$)}

\begin{figure}
\includegraphics[width=0.45\textwidth]{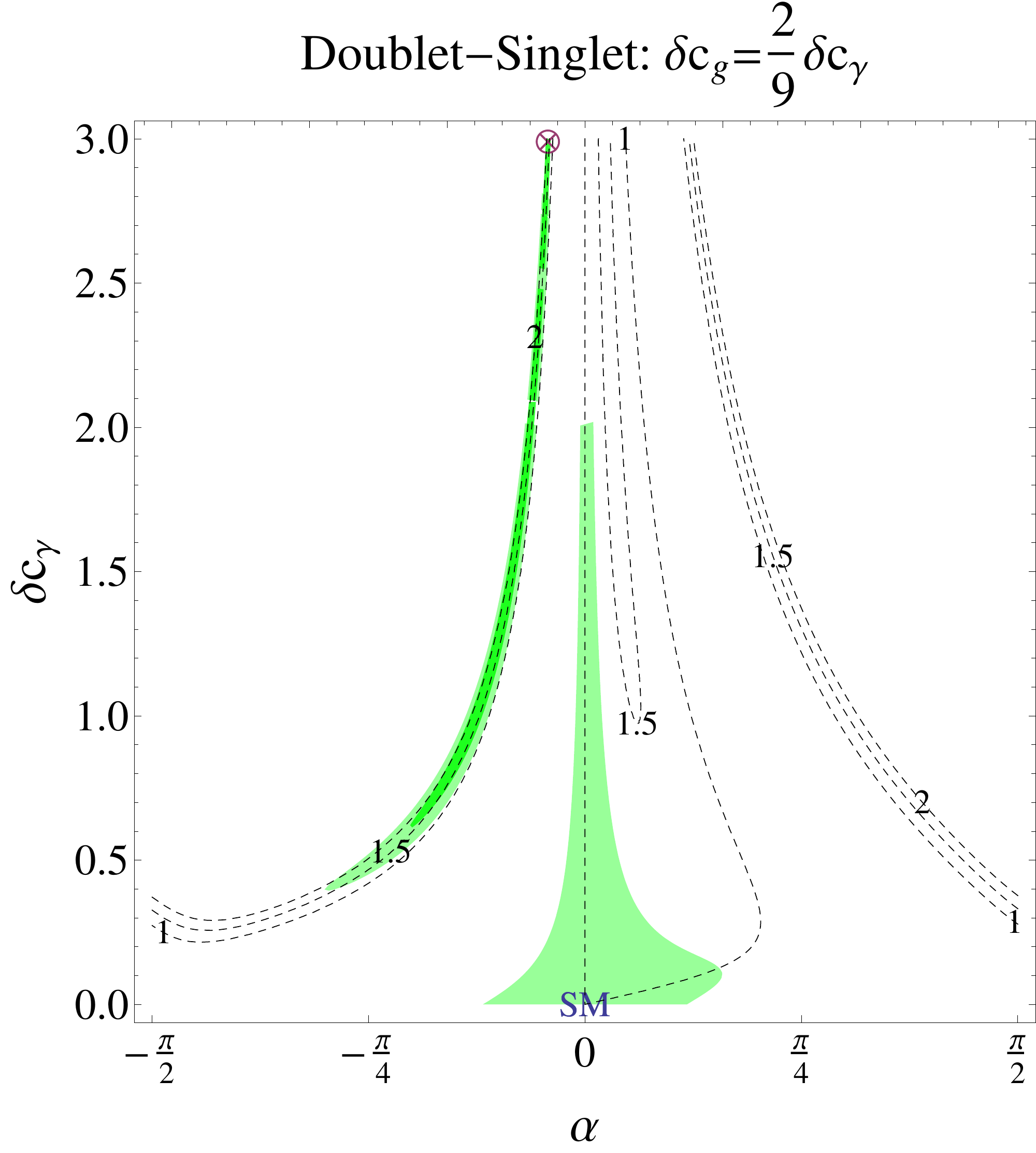}
\quad
\includegraphics[width=0.45\textwidth]{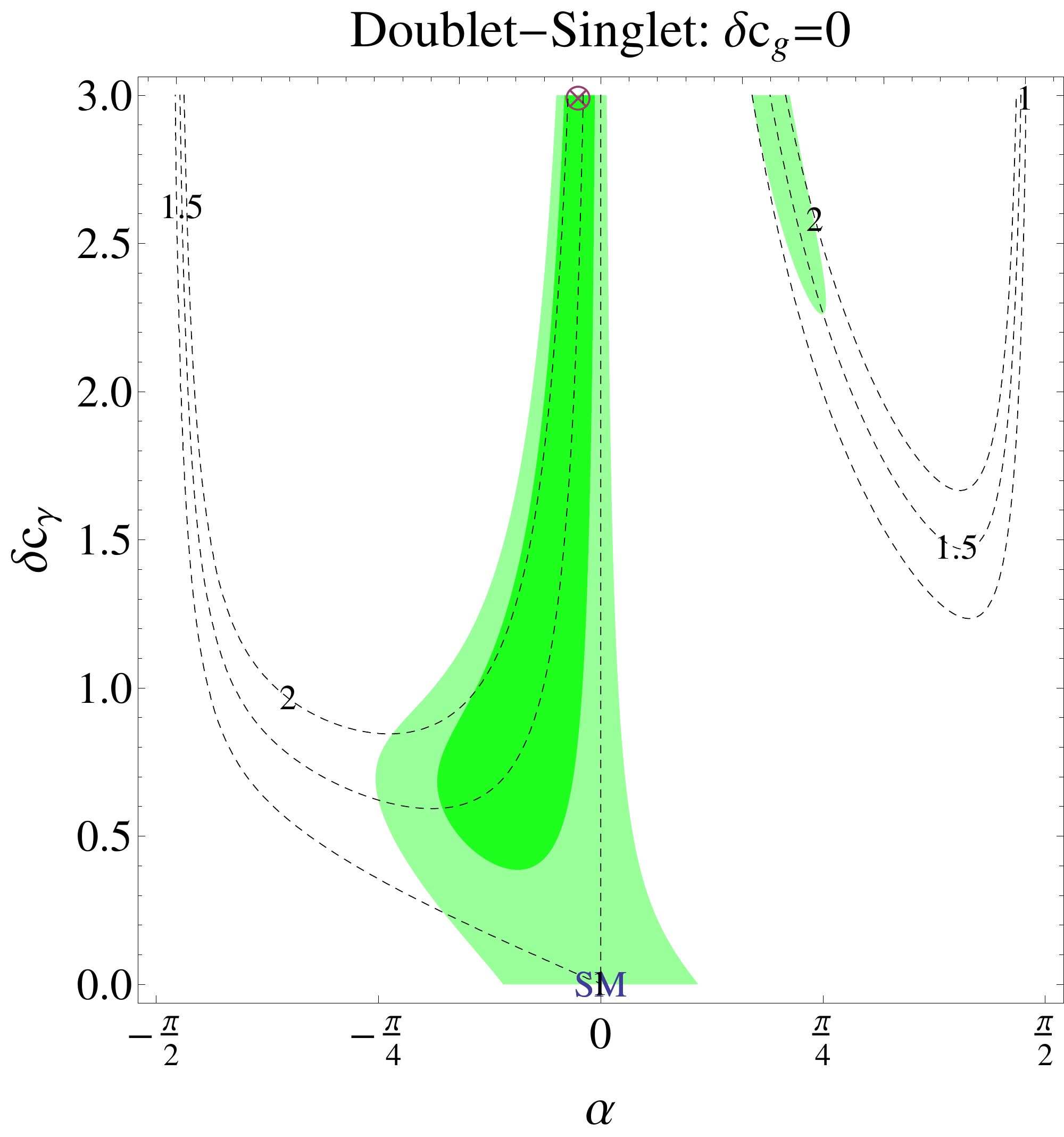}
\caption{\small
The best fit regions in the $\alpha$-$\delta_\gamma$ plane, assuming $m_h$=$125$ GeV and the singlet coupled to top-like (left) and colorless (right) vector-like fermions. Shown are $68\%$ (darker green) and $95\%$ CL (lighter green) regions.
The cross corresponds to the best fit point.  The dashed lines are contours of constant $R_{\gamma \gamma}$.
 \label{f.doubletsinglet}}
\end{figure}

The current CMS and ATLAS data point to a somewhat enhanced diphoton rate.   One simple scenario which may allow for such an enhancement is the case where the Higgs doublet mixes with a singlet, thereby affecting the lower-dimension parameters of the effective Lagrangian \eqref{eq:1}. More precisely, let us assume that  the SM Higgs doublet $H = (0,v+h_0)/\sqrt 2$ mixes with a real scalar, $\varphi$, which couples to new vector-like charged fermions $\psi_i$,
\begin{equation}
\mathcal{L} \supset -\frac{1}{2}m_\varphi^2 \varphi^2-\kappa \varphi |H|^2 - \sum_i M_i (1+ \frac{\lambda_i}{v}\varphi)\overline{\psi}_i\psi_i \, .
\end{equation}
The two mass-eigenstate Higgs scalars, $h$ and $H^0$,  mix with one another with an angle $\alpha$ defined by
\beq
h = h_0 \cos{\alpha} + \varphi \sin{\alpha} \,,
\qquad
H^0 = - h_0 \sin{\alpha} + \varphi \cos{\alpha} \,,
\eeq
and we assume $h$ to be the Higgs at $m_h \simeq125$ GeV.
The mixing angle suppresses the couplings of $h$ to $WW$, $ZZ$ and SM fermions by $\cos{\alpha}$  relative to the SM.
On the other hand, $h$ receives additional, possibly positive, contributions to the  couplings to $F_{\mu \nu}F^{\mu \nu}$ and $G_{\mu \nu}^a G^{a \mu \nu}$ from loops of the fermions $\psi_i$, resulting in
\begin{align}
c_{g} &= \,c_{g,{\rm SM}}\cos{\alpha} + \delta c_g \sin{\alpha} \,,\\
{c}_{\gamma} &= \, {c}_{\gamma,{\rm SM}}\cos{\alpha} +  \delta c_\gamma \sin{\alpha}\,.
\end{align}
Here the ``singlet" one-loop couplings $\delta c_g$ and $\delta c_\gamma$ are given by
\beq
\delta c_g = \sum_i 2 C_2(r_i)\lambda_i A_f(\tau_i)\,, \qquad
\delta c_\gamma = \sum_i \frac{1}{6} N_iQ_i^2 \lambda_i A_f(\tau_i) \, .
\eeq
The   fermion representations and couplings, $\lambda_i$, can always be chosen  to obtain any desired correction to $\delta c_g$ and $\delta c_\gamma$.

Mapping to the effective theory one finds,
\begin{eqnarray}
&&c_V = c_b = c_c = c_\tau = \cos{\alpha}\,,
\\
&&c_g = \cos{\alpha} A_f(\tau_t) +   \delta c_g \sin{\alpha}\,,
\\
&&c_\gamma = {2\over 9}\cos{\alpha} A_f(\tau_t) +  \delta c_\gamma \sin{\alpha}\,.
\end{eqnarray}
Accordingly, the double-singlet model has three free parameters: $\alpha$, $\delta c_g$, and $\delta c_\gamma$.
Two examples of motivated ansatzes are $\delta c_\gamma = (2/9) \delta c_g$ (when $\psi_i$ have quantum numbers of the top quark), and $\delta c_g = 0$ (when $\psi_i$ are color singlets).
The fits in the $\alpha$-$\delta c_\gamma$ plane for these two ansatzes are shown in \fref{doubletsinglet}\footnote{We thank Matthew McCullough for pointing out an error in the earlier version of this plot.}.
Only $\delta c_\gamma > 0$ is shown as  the region with $\delta c_\gamma < 0$ is equivalent upon $\alpha \to - \alpha$.
The best fit is  obtained for a negative mixing angle, which allows for an enhanced  Higgs diphoton rate, (as demonstrated by the constant contour lines).   On the other hand, the mixing angle is required to be sufficiently small in order to avoid strong suppression of the WW and ZZ rates.

Finally, we note that the doublet-singlet model predicts an additional resonance which may look  at colliders much like the Higgs field, but  with suppressed couplings.   While its mass remains a free parameter, the existence of such a state may therefore allow one to confirm or exclude this model in the near future.   We postpone further  details of this model for future work.

\subsection{The Doublet-Triplet Model (Enhanced $c_V$)}
\label{sec:doublet-triplet}

If the Higgs sector  contains triplets or higher representations under $SU(2)_W$ then the Higgs coupling to $W $and $Z$ bosons can be enhanced.
The Georgi-Machacek (GM) model  \cite{Georgi:1985nv,Gunion:1989ci,Low:2010jp} is one example that contains Higgs triplets, is renormalizable, and does not introduce large violations of the custodial symmetry.

The Higgs sector of the GM model contains the  usual Higgs doublet $H$ transforming as ${\bf 2}_{1/2}$ under $SU(2)_W\times U(1)_Y$, a real triplet $\phi$ transforming as ${\bf 3}_0$,
and a complex triplet $\Delta$ transforming as ${\bf 3}_1$. This field content forms irreducible representations of the approximate global symmetry group $G \equiv SU(2)_L \times SU(2)_R$:  the doublet can be put in  $\Phi$ forming the $(2,\bar 2)$ representation of $G$, while the triplets can be put in $\chi$ forming the $(3, \bar 3)$ of $G$.
The vev $v = 246$ GeV is distributed between the doublet and triplets as $\langle \Phi \rangle = v c_\beta I_{2 \times 2}$, $\langle \chi \rangle = v s_\beta I_{3 \times 3}$.
Since the $T$ parameter is protected by a built-in custodial symmetry, the triplet  vevs parameterized by  $s_\beta \equiv \sin \beta$ can be ${\mathcal O}(1)$ without conflicting phenomenology.
The vevs of $\Phi$ and $\chi$ break $G$ down to $SU(2)_V$ referred to as {\em custodial isospin}, under which $\Phi$ decomposes as singlet $H_{(2)}$ and triplet $G_{(2)}$, while $\chi$ decomposes as singlet $H_{(3)}$,  triplet $G_{(3)}$, and quintuplet  $Q$.
Of the custodial triplets there are only three physical states, $A^\pm,A^0$, while the  Goldstone bosons, $G$, get eaten by $W$ and $Z$. Here $A$ and $G$ are defined by, $G_{(2)}^a =  c_\beta G^a   - s_\beta A^a, $
 $G_{(3)}^a =  s_\beta G^a   + c_\beta A^a, $ where  $a=\pm,0$.   The isospin singlets also mix,
\beq
H_{(2)} =  c_\alpha h   - s_\alpha H\,,
\qquad
H_{(3)} =  s_\alpha h   + c_\alpha H\,,
\eeq
where the angle $\alpha$ depends on the details of the Higgs potential.
We identify the 125 GeV resonance with $h$.
Further  technical details about the model are postponed until Appendix~\ref{App:GM}.

The GM Higgs potential in the custodial limit contains 7 parameters: 2 masses and 5 quartic couplings, see \eqref{GM-potential}.
They can be related to 7 observables: the known VEV $v$, the 2 mixing angles $\alpha,\beta$, and the masses of the 2 isospin singlets, $m_h=125$~GeV  and $m_H$, the isospin triplet, $m_A$, and the isospin quintuplet, $m_Q$.
 In our convention the Higgs  is SM like for $\alpha = \beta = 0$.
The light Higgs phenomenology is thus determined by the 4 parameters $\alpha$, $\beta$, $m_Q$ and $m_A$, the former two affecting  the lower dimensional Higgs couplings, the latter affecting the dimension-5 couplings to gluons and photons.
Mapping to the effective theory in \eqref{eq:1} we find,
\beq
c_V =  c_\alpha c_\beta + \sqrt{8/3}  s_\alpha s_\beta, \qquad
c_b  =c_c=c_\tau=  {c_\alpha \over c_\beta} ,
\eeq
\beq
c_g   = c_b A_{f}(\tau_t),
\qquad
c_\gamma    =  {2 \over 9} c_b A_{f}(\tau_t) + { g_{h A^* A }  \over 24}  A_s(\tau_A) + {5  g_{h Q^* Q }  \over 24}  A_s(\tau_Q).
\eeq
Note that the coupling to $W$ and $Z$ can be enhanced over the SM value up to the maximal value of $c_V=\sqrt{8/3}$, which is a distinct feature of models that contain custodial quintuplets under the custodial symmetry.
The coupling to  the fermions can also be enhanced or suppressed compared to the SM Higgs, depending on $\alpha$ and $\beta$.
The $c_g$ and $c_\gamma$ differ from the SM value because of the modified Higgs-top coupling  while $c_\gamma$ also receives additional contributions from integrating out the charged Higgses $Q^{++}, Q^{+}$ and $A^+$. The couplings $g_{hA^*A}$ and $g_{hQ^* Q}$ are given in \eqref{ghAA} and \eqref{ghQQ}.

\begin{figure}[tb]
\begin{center}
\includegraphics[width=0.44\textwidth]{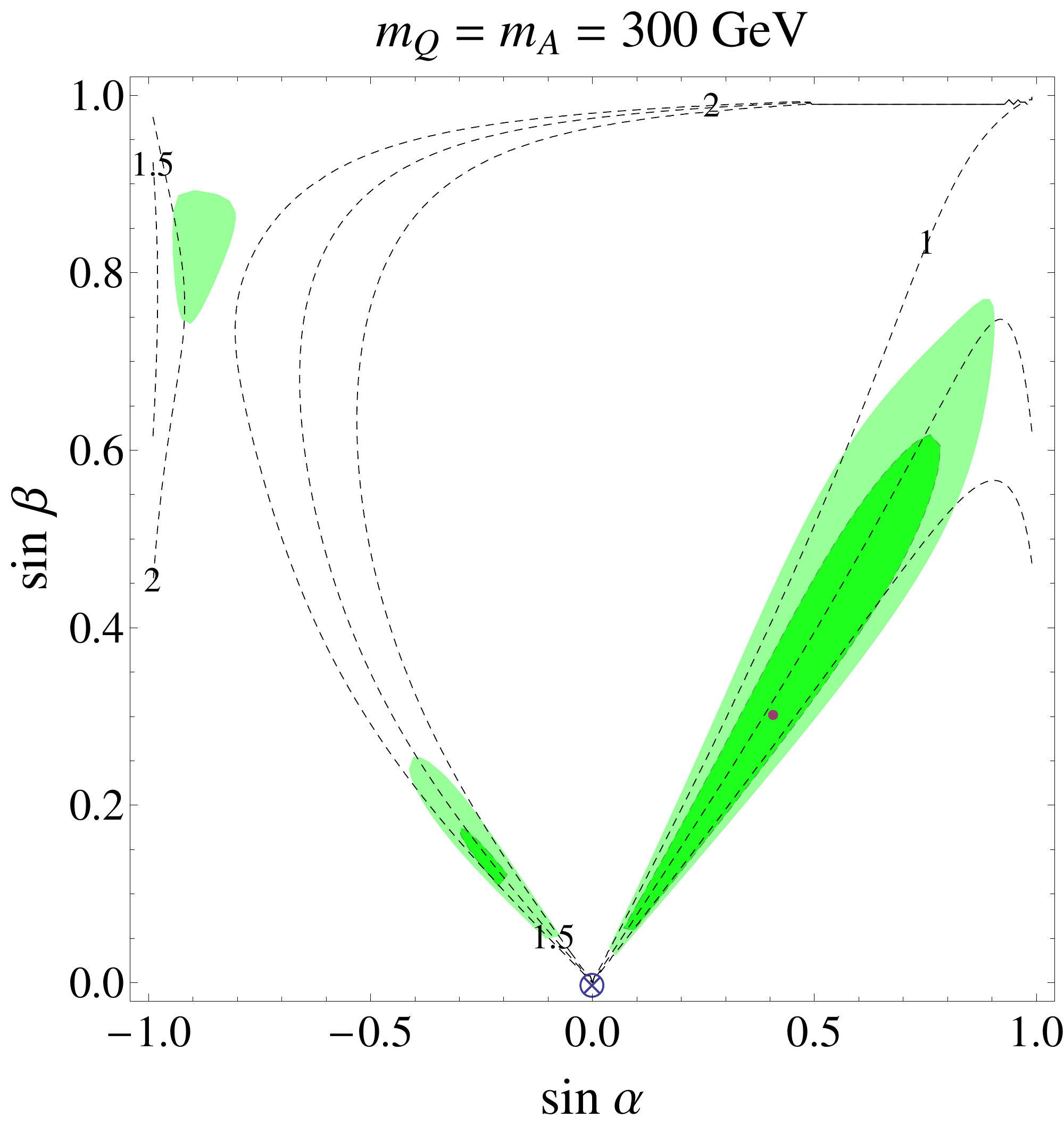}
\quad
\includegraphics[width=0.455\textwidth]{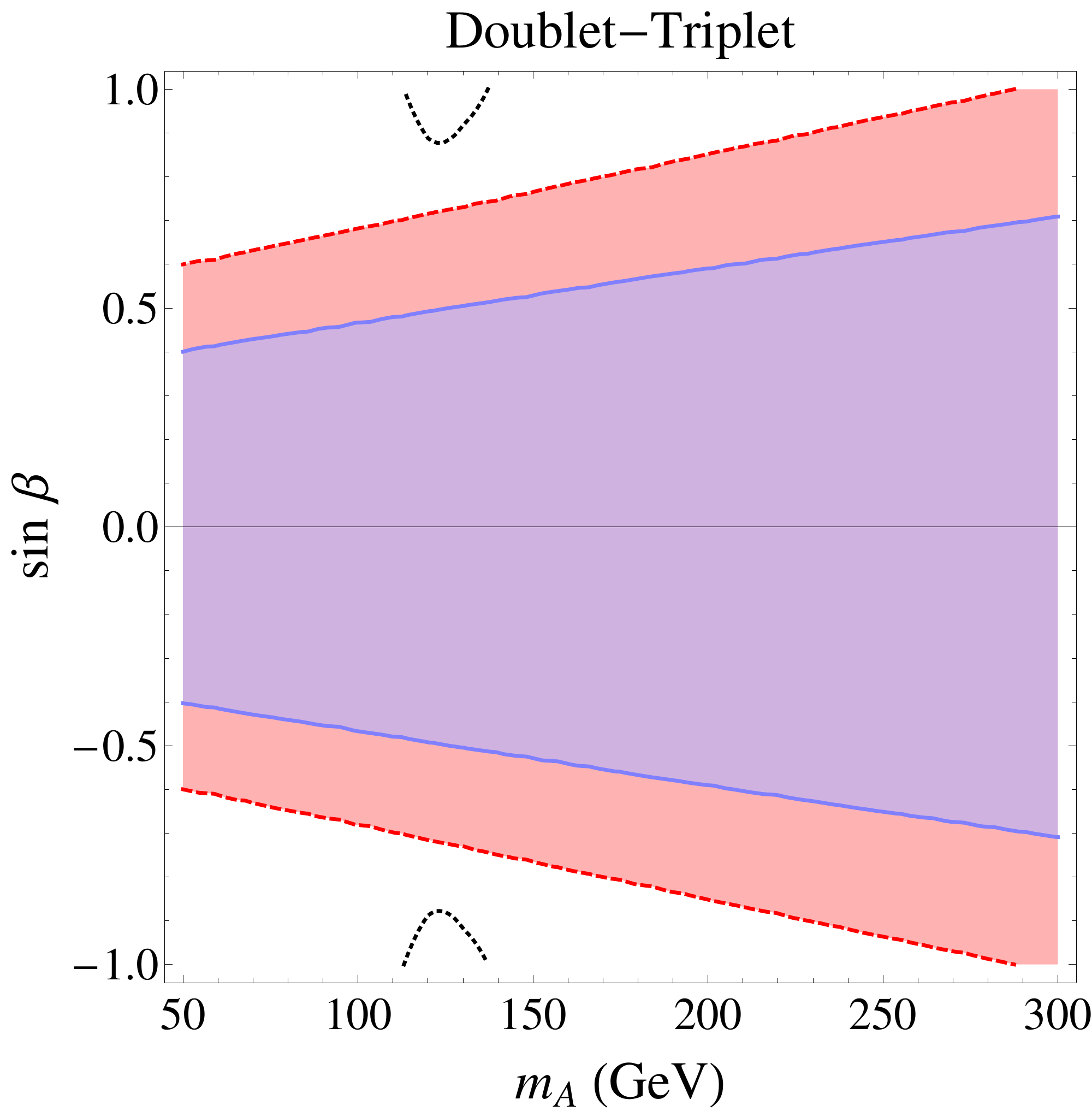}
\end{center}
\caption{\small
{\bf Left}: Best fit regions at $68\%$ (darker green) and $95\%$ CL (lighter green) in the $(\sin\alpha, \sin\beta)$ plane, setting $m_h = 125$ GeV
and $m_Q=m_A=300$ GeV. The SM point (cross)  compared to the  $\chi^2$ minimum (red dot) corresponds to  $\chi_{\rm SM}^2 - \chi_{\rm min}^2  = 4.4$.
The dashed lines show contours of constant $R_{\gamma \gamma}$.
{\bf Right}: The $95\%$ CL allowed regions for $\sin\beta$ and $m_A$ obeying constraints from $B_{s,d}-\bar B_{s,d}$ mixing (blue band with solid contour), from $b\to s\gamma$ (red band with dashed contour) and from $t\to b A^+$ decays (small regions inside dotted curves are excluded).
 \label{f.regsab100}}
\end{figure}

The best fit  regions in the $s_\alpha$-$s_\beta$ plane fixing $m_Q  = m_A = 300$ GeV are shown in the left panel of \fref{regsab100}.
The Higgs data are consistent with $s_\beta$ significantly larger than zero, that is with sizable triplet vevs.
The best fit regions are concentrated along $|s_\alpha| \sim s_\beta$ where the Higgs coupling to fermion is  near the SM value, but the coupling to $W$ and $Z$ is enhanced, leading to an enhanced diphoton rate.
Note that the best fit regions are well within ranges allowed by the other constraints (see also Ref.~\cite{Chang:2012gn} for a recent phenomenological study of the GM model).
The constraints on $m_A$ as a function os $s_\beta$ coming from $b\to s\gamma$, $B_{s,d}-\bar B_{s,d}$ and $t\to b A^+$ decays are shown in the right panel of \fref{regsab100}. We have also checked that the $m_Q=300$ GeV is not  constrained by doubly charged Higgs searches at CMS~\cite{HIG-12-005}, ATLAS~\cite{ATLAS-CONF-2011-127}, CDF~\cite{Aaltonen:2011rta} and DO \cite{Abazov:2011xx}, and from searches  for anomalously large production of multi-lepton final states \cite{Chatrchyan:2012ye} (similar sensitivity is expected from SUSY searches that require same sign leptons with missing jet and MET \cite{:2012th,Aad:2011cwa,:2012cwa}).

\section{Conclusions}
\label{sec:conclusions}

The recent discovery of a Higgs boson at 125 GeV has important consequence for any new physics scenario.
 In this paper, we studied those implications using the Higgs searches reported by ATLAS, CMS and the Tevatron in currently the most sensitive channels.
We derived the constraints on the parameters of the effective Lagrangian describing, in a very general fashion, the leading order interactions of the Higgs particle with matter.
Overall, the point in the parameter space corresponding to the SM Higgs boson is well consistent, within 74\% CL, with the Higgs data.
The data show a preference for models leading to an increased rate in the diphoton search channels, but at this point this is not statistically significant.
We note that the current data are still consistent with a sizable invisible branching fraction of the Higgs:  
${\rm Br}_{inv} \lesssim 20\%$ if the Higgs production rate is the one predicted by the SM, and ${\rm Br}_{inv} \lesssim 65\%$, if new large contributions to the effective Higgs coupling to gluons and photons are present.

As a second step, we mapped a number of new physics models affecting Higgs phenomenology on to the effective Lagrangian, placing constraints on the relevant couplings.
In particular, we study the simplified version of the MSSM, composite Higgs models, dilaton, a two-Higgs doublet model, a doublet-singlet model and a doublet-triplet model.
In each case we identified the region of the parameter space that can improve the quality of the fit by enhancing the rate in the diphoton channel.
If that feature persists with more data, indicating a departure from the Standard Model, new physics models would require special structure and additional charged particles, likely within collider reach, in order to explain the data.

{\bf Note added:} While this work was being completed the analyses \cite{Low:2012rj} appeared that overlap to some extent with our results.

\vspace{1cm}

\section*{Acknowledgements}

We thank Jeremy Mardon for early collaboration.   We also thank  Gideon Bella, Jamison Galloway, Matthew McCullough, and Alberto Romagnoni  for useful discussions.
The work of DC, EK and TV is supported in part by a grant from the Israel Science Foundation.   The work of TV is further supported in part by the US-Israel Binational Science Foundation and the EU-FP7 Marie Curie, CIG fellowship. JZ was supported in part by the U.S.\ National Science Foundation under CAREER
Award PHYÐ1151392.

\appendix
\section{Further details on the Georgi-Machacek model}
\label{App:GM}

We first introduce some notation.
The $J=1/2$ SU(2) generators are $T^a = \sigma^a/2$,  where $\sigma^a$ are the usual Pauli matrices,
while  $T_1^a$ are  the SU(2) $J =1$ generators in charge eigenstate basis,
\beq\label{e.su2gen_charge}
T_{1}^1 = \frac{1}{ \sqrt 2}
\begin{pmatrix}
0 & -1 & 0 \\ -1 & 0 & 1\\ 0 & 1 & 0
\end{pmatrix},
\quad
T_1^2 =   \frac{1}{ \sqrt 2}
\begin{pmatrix}
  0 & i & 0 \\ -i & 0 & -i\\ 0 & i & 0
  \end{pmatrix},
\quad
T_1^3 =
\begin{pmatrix}  1 & 0 & 0 \\ 0 & 0 & 0 \\ 0 & 0 & -1
\end{pmatrix}.
\eeq
We  also defined the matrix  $P_3$
\beq
\label{e.su2gen_p3}
P_3 =  \left ( \ba{ccc}  0 & 0 & 1 \\ 0 & 1 & 0 \\ 1 & 0 & 0 \ea \right )
\eeq
that satisfies $P_3 (T^a)^* = - T^a P_3$. Thus, if $\psi$ transforms as a triplet, then $\ti \psi \equiv P_3 \psi^*$ transforms as a triplet too.

In the GM model, the electroweak doublet $H$ and the triplets $\phi$, $\Delta$ can be collected into $(2,\bar 2)$ and $(3,\bar 3)$ representation under the global custodial $G = SU(2)_L \times SU(2)_R$:
\beq
\Phi = \left (\ti H  \ H \right ),
\qquad
\chi = \left ( \ti \Delta \ \phi \ \Delta \right).
\eeq
Here $\ti H \equiv i \sigma^2 H^*$ and $\ti \Delta \equiv P_3 \Delta^*$. 
In this notation the $U(1)_Y$ is realized as the $T^3$ generator of $SU(2)_R$.
The vevs of $\Phi$ and $\chi$ break $G$ down to the diagonal $SU(2)$ referred to as the custodial isospin,
 under which $\Phi$ decomposes as singlet $H_{(2)}$ and triplet $G_{(2)}$, while
 $\chi$ decomposes as singlet $H_{(3)}$,  triplet $G_{(3)}$, and quintuplet  $Q$.
These fields  are embedded into the doublet and triplets as,
\beq
H = \bvec i G_{(2)}^+ \\ \tfrac{1}{\sqrt2}(v c_\beta + H_{(2)} - i G_{(2)}^0) \evec ,
\eeq
\beq
\phi = \bvec
{1 \over \sqrt{2}} ( Q^+  - i G_{(3)}^+)
\\
{v s_\beta \over 2 \sqrt 2}   + \sqrt{1 \over 3} H_{(3)} -   \sqrt{2 \over 3} Q^0
\\
 {1 \over \sqrt{2}} (Q^- +  i G_{(3)}^-)
\evec, \quad
\Delta = \bvec
Q^{++}
\\
 {1 \over \sqrt{2}} (Q^+  + i  G_{(3)}^+)
\\
{v s_\beta \over 2 \sqrt 2}  +   \sqrt{1 \over 3} H_{(3)} +  \sqrt{1 \over 6} Q^0   - \frac{1}{\sqrt2}i G_{(3)}^0
\evec .
\eeq

The kinetic terms in the Higgs sector are given by
\beq
\label{e.kin}
\cl_{kin} =  {1 \over 2} {\rm Tr} \left [ D_\mu \Phi^\dagger D_\mu \Phi \right ]    +  {1 \over 2} {\rm Tr} \left [ D_\mu \chi^\dagger D_\mu \chi \right ],
\eeq
with the covariant derivatives
\begin{align}\label{GM:covariant1}
D_\mu \Phi &=  \pa_\mu \Phi -  i g {\sigma^a \over 2} L_\mu^a \Phi + i g' B_\mu \Phi {\sigma^3 \over 2}, \\
 D_\mu \chi &=  \pa_\mu \chi -  i g T_1^a  L_\mu^a \chi + i g' B_\mu \chi T_1^3, \label{GM:covariant2}
\end{align}
One can check that \eref{kin} leads to the kinetic terms for the scalars that are diagonal and canonically normalized, and it also leads to the usual SM gauge  boson masses with $W^\pm$ and $Z$ kinetically mixing with Goldstone bosons $G^\pm$ and $G^0$.
The most general custodially invariant and renormalizable potential is  \cite{Gunion:1989ci}
\beq
\begin{split}\label{GM-potential}
V =&      {m_\Phi^2 \over 2} {\rm Tr} \left [ \Phi^\dagger \Phi \right ]    +  {m_\chi^2 \over 2} {\rm Tr} \left [ \chi^\dagger \chi \right ]
 + \lambda_1 \left ({\rm Tr} \left [\Phi^\dagger \Phi \right ]  \right )^2    +  \lambda_2 {\rm Tr} \left [  \Phi^\dagger  \Phi \right ]  \tr [\chi^\dagger \chi] \\
& +  \lambda_3  \tr [\chi^\dagger \chi \chi^\dagger \chi] +  \lambda_4 \left(  \tr [\chi^\dagger \chi] \right )^2
- \lambda_5 {\rm Tr} \left [ \Phi^\dagger \sigma^a \Phi  \sigma^b \right ]  \tr [\chi^\dagger T_1^a \chi T_1^b].
\end{split}
\eeq
The quark Yukawa terms only involve the Higgs electroweak doublet,
 \beq\label{GM:Yukawa}
\cl =  - {y_q \over \sqrt 2} \left (v c_\beta + H_{(2)} \right )   \bar q_L q_R      \to  -  m_q  \left (1  + {c_\alpha \over c_\beta} {h \over v}  - {s_\alpha \over c_\beta} {H \over v}   \right )    \bar q_L q_R,
 \eeq
where  $m_q =  y_q v c_\beta/\sqrt 2$.
In the lepton sector there the Yukawa couplings to both  the Higgs doublet $H$ and the triplet $\Delta$ are possible.
The coupling of the doublet is of the same form as for the quarks, $\cl =  - y_\ell H  L e^c + \hc$,
 while the coupling to the triplet, $ \cl= -\lambda_\ell \Delta^a  L \sigma^a L + \hc$, also contributes to  Majorana neutrino masses as soon as  $s_\beta\ne 0$.
Given the smallness of the neutrino masses the Yukawa coupling constants $\lambda_\ell$ are exceedingly small for realistic $s_\beta$, and can be neglected for our purposes.

The $h\to 2\gamma$ decay is affected  by the charged component of the custodial isospin triplets and quadruplets running in the loop. Keeping only the  trilinear couplings of $h$ to the triplet and quintuplet in the potential \eqref{GM-potential} one gets
 \beq
  \cl  = -  g_{h A^* A }{m_A^2} h \left ( 2 A^+ A^- + A_0^2 \right )-  g_{h Q^* Q }  {m_Q^2}  h \left ( 2Q^{++} Q^{--} + 2 Q^+ Q^- + Q_0^2 \right ),
 \eeq
 where
 \begin{align}\label{ghAA}
 g_{h A^* A } &= \big( c_\alpha c_\beta + \sqrt {8/3} s_\alpha s_\beta \big)
 + \left(m_h^2/ m_A^2\right)   \big(2 \sqrt 6 c_\beta^3 s_\alpha +  3 c_\alpha s_\beta^3\big)/\left( 6 c_\beta s_\beta  \right) ,
\\
\label{ghQQ}
 g_{h Q^* Q } &=
\sqrt{2/3}\big(2 + {m_h^2/ m_Q^2} \big){s_\alpha}/{ s_\beta }
 + \left(m_A^2 / m_Q^2\right)  c_\beta (-2 \sqrt 6  c_\beta s_\alpha + 3 c_\alpha s_\beta)  / s_\beta.
 \end{align}

 Let us now review the constraints on masses  of the extra scalars, $H$, $A$ and $Q$. The heavy Higgs is constrained by direct searches at the LHC for the SM Higgs decaying to $WW$ and $ZZ$. The bounds are avoided for $m_H>600$ GeV, while for smaller values $s_\beta$ is constrained \cite{Chang:2012gn}. The charged $A^+$  boson couples to fermions through the Yukawas  which in the mass eigenstate basis are
\beq
{\cal L}_{\rm int}=i (\bar u_L V_{CKM} Y_d d_R) s_\beta A^++i (\bar u_R Y_U V_{CKM}^\dagger d_L) s_\beta A^+ +h.c.,
\eeq
where $Y_{u,d}= {\rm diag}(m_{u,d})\sqrt{2}/v$ and $V_{CKM}$ is the CKM matrix. The FCNCs are generated at 1-loop. The most constraining are $b\to s\gamma$ and $B_{d,s}-\bar B_{d,s}$ mixing with  $A^+$ and top in the loop. To obtain the bounds we use LO matching  onto effective weak Hamiltonian at $\mu_W$. For $b\to s\gamma$ we use the LO equations  (53-56) in \cite{Ciuchini:1997xe} with the replacement $A_{u,d}=\pm s_\beta/(v\sqrt{2G_F})$ (the NLO results are also available \cite{Ciuchini:1997xe,Borzumati:1998tg}), the parametrization in Eqs. (46-49) of \cite{Kagan:1998ym} for the effect on $b\to s\gamma$, the experimental value of $Br(B\to X_s\gamma)=(3.55\pm0.26)\cdot 10^{-4}$ \cite{Asner:2010qj} and the SM prediction $Br(B\to X_s\gamma)_{\rm SM}=(3.15\pm0.23)\cdot 10^{-4}$ \cite{Misiak:2006zs} (for the estimate of the irreducible error, see \cite{Benzke:2010js}).
The resulting bounds on $s_\beta$ and $m_A$ are shown in Fig. \ref{f.regsab100} as a red band. For constraints from $B_{d,s}-\bar B_{d,s}$ mixing we used the LO matching expression in Eq. (54-56) of~\cite{Urban:1997gw} with the replacement $1/\tan\beta\to s_\beta$. The ratio of NP to SM contributions $h_{d,s}=M_{12}^{\rm NP}/M_{12}^{\rm SM}=(2 S_{WH}+S_{HH} )/S_{WW}$ is the same for $B_{d}-\bar B_d$ and $B_s-\bar B_s$ systems since the Georgi-Machacek model is an example of an MFV extension of the SM. We therefore use the results of the fit where MFV constraints are imposed on the mixing amplitudes, $h_{d,s}=-0.08^{+0.12}_{-0.039}$ \cite{Lenz:2012az} (this is Scenario II in the terminology of \cite{Lenz:2012az}) leading to constraints on $s_\beta$ and $m_A$ shown in Fig. \ref{f.regsab100} as a blue band. Note that in Georgi-Machacek model $h_{d,s}$ is always positive. Since the central value of $h_{d,s}$ is negative experimentally (but well within one sigma range) this makes the constraints slightly more stringent.

Finally, the direct searches relevant for $Q$ constraints are direct searches for $H^{++}$, $H^+\to \tau \nu$ and SUSY searches with leptons and MET. In the doubly charged Higgs searches from CMS~\cite{HIG-12-005}, ATLAS~\cite{ATLAS-CONF-2011-127}, CDF~\cite{Aaltonen:2011rta} and DO \cite{Abazov:2011xx} the dominant decay modes were assumed to be $Q^{++}\to \ell^+\ell^+$ and $Q^+\to \ell^+\nu$.  In Georgi-Machacek models the dominant decays are not into leptons but into gauge bosons, which make the searches not sensitive to $Q$ at present. Similarly we have checked that the search for anomalously large  production of multi-lepton final states is not sensitive to $m_Q\sim 100$~GeV \cite{Chatrchyan:2012ye} and similar (in)sensitivity is expected from other SUSY searches \cite{:2012th,Aad:2011cwa,:2012cwa}.

More constraining are the charged Higgs searches in the $H^+\to \tau\nu$ channel, both in CMS \cite{:2012cw} and ATLAS \cite{Aad:2012tj}. These are only relevant, if $A^+$ is lighter than $t$. In this case the branching ratio for $Br(t\to bA^+)Br(A^+\to \tau^+\nu)$ is bounded to be below between $4\%$ for $m_A=90$ GeV and $1\%$ for $m_A=160$ GeV at 95 C.L.  \cite{Aad:2012tj} (the limits are comparable for CMS). The branching ratio $Br(t\to A^+ b)\simeq s_\beta^2$ in Georgi-Machacek model, if we neglect the effect of phase space for clarity. On the other hand $Br(A^+\to \tau^+\nu_\tau):Br(A^+\to c\bar s):Br(A^+\to c\bar b)\simeq m_\tau^2:m_c^2:|V_{cb}|^2 m_t^2=0.06:0.03:0.91$, with masses evaluated at the scale $\mu\simeq m_Z$. This means that $A^+\to \tau^+\nu_\tau$ is a subdominant decay mode with only $\sim 6\%$ branching ratio. The ATLAS search thus excludes at 95\%CL  a small region of $(m_A, s_\beta)$ parameter space around $m_A\sim 120$ GeV, see  \fref{regsab100} (CMS is just about to become constraining).

\end{document}